\documentclass[english]{revtex4-1}
\usepackage[T1]{fontenc}
\usepackage[latin9]{inputenc}
\usepackage{mathtools}
\usepackage{amsmath}
\usepackage{amssymb}
\usepackage{graphicx}
\usepackage{esint}
\usepackage{babel}
\begin{document}
\title{Dressed Quantum Trajectories: Novel Approach to the non-Markovian
Dynamics of Open Quantum Systems on a Wide Time Scale}
\author{Evgeny A. Polyakov$^{1}$, Alexey N. Rubtsov$^{1,2}$}
\address{$^{1}$Russian Quantum Center, 100 Nonaya St., Skolkovo, Moscow 143025,
Russia}
\address{$^{2}$Department of Physics, Lomonosov Moskow State University, Leninskie
gory 1, 119991 Moscow, Russia}
\begin{abstract}
A new approach to the theory and simulation of the non-Markovian dynamics
of open quantum systems is presented. It is based on identification
of a parameter which is uniformly small on wide time intervals: the
occupation of the virtual cloud of quanta. By ``virtual'' we denote
those bath excitations which were emitted by the system, but eventually
will be reabsorbed before any measurement of the bath state. A favourable
property of the virtual cloud is that the number of its quanta is
expected to saturate on long times, since physically this cloud is
a (retarded) polarization of the bath around the system. Therefore,
the joint state of open system and of virtual cloud (the dressed state)
can be accurately represented in a truncated basis of Fock states,
on a wide time scale. At the same time, there can be arbitrarily large
number of observable quanta, especially if the open system is under
driving. However, by employing a Monte Carlo sampling of the measurement
outcomes of the bath, we can simulate the dynamics of the observable
quantum field. In this work we consider the measurement with respect
to the coherent states, which yields the Husimi function as the positive
(quasi)probability distribution of the outcomes. The evolution of
dressed state which corresponds to a particular fixed outcome is called
the dressed quantum trajectory. Therefore, the Monte Carlo sampling
of these trajectories yields a stochastic simulation method with promising
convergence properties on wide time scales. 
\end{abstract}
\maketitle

\section{INTRODUCTION}

The model of a finite quantum system coupled to an inifinite harmonic
bath (the open quantum system, abbreviated as OQS) is of great importance
for numerous branches of quantum physics. It is this model on which
the concepts of measurement and decoherence are worked out \citep{Zurek2003,Schlosshauer2004},
from the fundamental questions of the emergence of classical world
\citep{Blume-Kohout2008,Riedel2011,Korbicz2017,Mironowicz2017,Knott2018},
to the modern protocols of adaptive quantum control \citep{Brandes2010,Kiesslich2012,Gough2014,Brandes2016,Luo2016,Wagner2016}
and information processing \citep{Beige2000,Verstraete2009,Zanardi2016,Kapit2018}.
In physical chemistry the concept of OQS is applied to describe the
transfer of phononic or electronic energy in the molecular complexes
\citep{deVega2017}. Finally, in the condensed matter physics, a special
type of OQS, the Anderson impurity model within a self-consistent
environment, is the central component of the dynamical mean-field
theory (DMFT) calculations \citep{Pruschke1995,Georges1996,Freericks2006,Aoki2014}.
Theoretical and experimental advances in all these branches continuously
raise challenging questions about how to properly characterize the
physical state of an open system, and how to simulate its dynamics
in various regimes. 

In the Markovian regime, when the environment recovers instantly after
the OQS disturbances, it is fairly well understood how to characterize
and propagate the state of open system: its state is represented either
by a reduced density matrix, which is governed by a master equation
\citep{Breuer2016}; or by a wavefunction, which is governed by a
stochastic Schrodinger equation \citep{Wiseman1993,Wiseman1996,Percival1999,Warszawski2003,Warszawski2003a,Oxtoby2005,Tilloy2015,Bauer2015,Breuer2016,Daley2014,Zhang2017}.
Closely related to these methods is the input-output formalism \citep{Gardiner1985,Gardiner1987,Gardiner2004},
which allows one to take into account the properties of the incident
and scattered excitations of the bath. 

On the contrary, the non-Markovian regime \citep{Breuer2016,deVega2017,Li2018},
when the bath has the memory of the OQS disturbances, is much less
understood \citep{Jack1999,Xu2018}. One of the main challenges of
the non-Markovain regime is the entanglement: as the time goes, the
OQS and the bath continuously exchange the quanta between themselves,
and the dimension of the joint entangled state grows with time, which
also leads to the combinatorially-increasing complexity of description
and simulation. In the Markovian case the situation was rather simple:
at the very moment of absorbtion and emission, the OQS state experiences
a sudden jump. Right after that, the bath forget the disturbance \citep{Breuer2011}.
However, in the non-Markovian regime, the emitted quanta maintain
the entanglement with the OQS for a certain period of time. In the
literature, there is the understanding that OQS cannot be entangled
to the whole bath all the time: OQS should eventually forget about
the emitted quanta \citep{Diosi2012}. But the question of how to
partition the bath $B$ into the two parts, $B=M+D$ , the entangled
memory part $M$, and the ``detector'' part $D$ of irreversible
emitted and forgotten quanta, still has no consise and transparent
answer, albeit there are investigations and proposals of possible
decompositions \citep{Garraway1996,Mazzola2009,Diosi2012,Arrigoni2013,Budini2013,Dorda2014}. 

More progress is achieved in the numerical simulation algorithms for
the non-Markovian quantum dynamics. For example, the matrix product
states approach \citep{Strathearn2018} was recently proposed to efficiently
calculate the discretized path integral with the influence functional
for the bath (the so-called Quasi-Adiabatic Path Integral \citep{Makarov1994,Makri1995,Makri1995a,Makri1996}
for OQS). However, it involves the truncation of the long-range tails
of the bath memory function after a certain time $\tau_{\textrm{cut}}$.
In the strong coupling regimes, such a truncation always distorts
the large-time asymptotic behaviour of the observables \citep{Strathearn2018}.
Increasing $\tau_{\textrm{cut}}$ leads to exponential increase of
the computational complexity.

Another promising family of algorithms are the stochastic simulation
techinques. In these algorithms, the interaction of the bath with
OQS is splitted into the two parts: one is represented by a stochastic
coloured noise (whose correlator reproduces the bath memory function);
the other part is solved deterministically. These algorithms include
the non-Markovian quantum state diffusion (NMQSD) approach \citep{Diosi1998},
the hierarchy of pure states (HOPS) \citep{Suess2014,Hartmann2017},
and the hierarchical approach based on stochastic decoupling \citep{Shao2004,Yan2004,Shao2008,Yan2016}.
These algorithms intend to simulate the real-time dynamics by a Monte-Carlo
sampling. However, in order to avoid the sign problem (since currently
every full real-time Monte-Carlo simulation suffers from it), these
algorithms have to simulate a certain part of quantum dynamics by
a deterministic scheme. Again, the deterministic part involves the
trunctation of the bath memory function after a certain time $\tau_{\textrm{cut}}$,
and increasing $\tau_{\textrm{cut}}$ leads to the exponential increase
of complexity. Another limiting factor of these methods is the coupling
strength between the bath and the OQS. The deterministic part is based
on a certain truncated hierarchy of OQS-bath correlation functions,
and the truncation level is increased with the coupling strength. 

In summary, a successful numerically exact solver for the non-Markovian
open quantum system should contain the following three ingredients:
(1) a physically motivated division into the stochastic and determenistic
part, (2) a truncation of the determenistic part suitable for a bath
with the long-range memory tails, and (3) a way to deal with the strong
coupling regime. We address the problem (1) in this paper, and simultaneously
propose a solution of the problem (2) in another manuscript \citep{Polyakov2018b}. 

Our approach to the non-Markovian quantum dynamics is based on an
identification of a small parameter on wide time scales. There are
conventional techniques which are based on a \textit{short-time} small
parameters e.g. any perturbative expansion in coupling. But these
techniques squickly loose their quantitative and qualitative accuracy
as the time scale is increased. We propose to consider the population
of the cloud of virtual bath quanta as a small parameter on wide time
scales. What we mean is the following. When the open quantum system
interacts with the surrounding bath, it emits and absorbs quanta (bath
excitations). We classify the quanta into the following types: virtual
and output \citep{Polyakov2017b,Polyakov2017a}. By ``virtual''
we denote those excitations which were emitted by OQS, but eventually
will be reabsorbed before any measurement of the bath state. At the
same time, the output quanta are those which will survive up to the
bath measurement time moment. We suggest that the joint state ``OQS
+ virtual quanta'', also called the \textit{dressed OQS state}, is
the appropriate characterisation of the physical state of OQS in a
non-Markovian regime. We take the analogies from the other branches
of physics: the physical state of electron which interacts with electromagnetic
field, is the ``bare'' electron dressed by a cloud of virtual photons;
in solid state, the electron is always dressed by a cloud of virtual
phonons. From these analogies, one may guess that the number of virtual
quanta shoud be bounded even at long simulation times and in presence
of driving. On a physical grounds, we expect it to be small for moderate
coupling. That is very favourable property for numerical methods,
since we will represent the dressed OQS state in a truncated Fock
basis for virtual quanta. 

At the same time, the output quantum field may contain arbitrarily
high number of quanta at large time scales, especially if there is
a driving, or the coupling contains the terms beyond the rotating-wave
approximation (RWA). Fortunately, the physical properties of the output
field provide us with the escape from the ensuing complexity: the
output field is always measured. We can employ a Monte Carlo sampling
of the measurement results in order to simulate the dynamics of the
output field. The most convenient basis for the measurement is provided
by the coherent states. Then, the quantum output fields are replaced
by classical fluctuating fields, resulting in a \textit{dressed }non-Markovian
quantum state diffusion (shortly dressed NMQSD) equation of motion. 

The proposed approach is founded on many partial results in the literature.
In particular, the idea of how to sample stochastically the observable
state of the bath was first introduced in the non-Markovian quantum
state diffusion approach (NMQSD) \citep{Diosi1998}. Another idea,
that the correct non-Markovian description of the physical state of
OQS should involve additional degrees of freedom (besides the reduced
density matrix), is repeatedly expressed in literature in various
contexts \citep{Tanimura1990,Garraway1996,Breuer2007,Mazzola2009,Diosi2012,Arrigoni2013,Budini2013,Dorda2014,Suess2014}.
The distinction between the two types of photons (the bath quanta)
- those which will be eventually absorbed by the detector, and those
which will be reabsorbed by the OQS - was first discussed in the works
of M. W. Jack \textit{et al }\citep{Jack1999}\textit{. }

In section \ref{subsec:Model-of-Open} we introduce the model of open
quantum system and then in \ref{subsec:Dressed-non-Markovian-quantum-1}
introduce the notion of a dressed OQS state corresponding to a fixed
measurement outcome for the bath. The master equation for the probability
distribution of the measurement outcomes, the Husimi function, is
derived in section \ref{subsec:Master-equation-for}. The stochastic
interpretation of this master equation is provided in terms of the
dressed quantum trajectories, as shown in section \ref{subsec:Husimi-dressed-quantum}.
Summary of the simulation algorithm and the expample calculation are
provided in sections \ref{subsec:Summary-of-the} and \ref{subsec:Example-calculation}.
In section \ref{sec:THE-PROBLEM-OF} we show that the problem of memory
tails for a driven quantum system arises at large times even at small
couplings. This problem is dealt with in the companion paper. We conclude
in section \ref{sec:CONCULSION}. In appendix \ref{sec:NORMALIZED-HUSIMI-QUANTUM}
we provide the details of derivation of the Husimi master equation.

\section{DRESSED NON-MARKOVIAN QUANTUM TRAJECTORIES}

In this section, we present our approach to the description and simulation
of open quantum system dynamics.

\subsection{Model of Open System\label{subsec:Model-of-Open}}

We consider a system which is linearly coupled to the bath of harmonic
oscillators. The Hamiltonian is
\begin{equation}
\widehat{H}=\widehat{H}_{\textrm{s}}+\hat{s}\widehat{b}^{\dagger}+\hat{s}^{\dagger}\widehat{b}+\widehat{H}_{\textrm{b}},\label{eq:total_hamiltonian_schrodinger_picture}
\end{equation}
where $\widehat{H}_{\textrm{s}}$ is the OQS, $\widehat{H}_{\textrm{b}}$
is the bath
\begin{equation}
\widehat{H}_{\textrm{b}}=\intop_{0}^{+\infty}d\omega\omega\widehat{a}^{\dagger}\left(\omega\right)\widehat{a}\left(\omega\right).\label{eq:free_bath_hamiltonian}
\end{equation}
The bosonic annihilation operators $\widehat{a}\left(\omega\right)$
obey to the canonical commutation relations 
\begin{equation}
\left[\widehat{a}\left(\omega\right),\widehat{a}^{\dagger}\left(\omega^{\prime}\right)\right]=\delta\left(\omega-\omega^{\prime}\right).\label{eq:canonical_commutation_relations}
\end{equation}
The coupling is through the operator $\widehat{s}$ which is in the
system's Hilbert space, and through the operator $\widehat{b}$ which
is in the Hilbert space of bath, 
\begin{equation}
\widehat{b}=\intop_{0}^{+\infty}d\omega c\left(\omega\right)\widehat{a}\left(\omega\right).\label{eq:coupling_to_the_bath}
\end{equation}
In our representation, the frequency dependence of the density-of-states
is transfered to the coupling coefficient $c\left(\omega\right)$.
In the interaction picture with respect to the free bath, the Hamiltonain
(\ref{eq:total_hamiltonian_schrodinger_picture}) becomes
\begin{equation}
\widehat{H}\left(\tau\right)=\widehat{H}_{\textrm{s}}+\hat{s}\widehat{b}^{\dagger}\left(\tau\right)+\hat{s}^{\dagger}\widehat{b}\left(\tau\right),\label{eq:total_hamiltonian_interaction_picture}
\end{equation}
with
\begin{equation}
\widehat{b}\left(\tau\right)=\intop_{0}^{+\infty}d\omega c\left(\omega\right)\widehat{a}\left(\omega\right)\exp\left(-i\omega\tau\right).
\end{equation}
 Properties of the bath are defined in terms of its memory function
\begin{equation}
M\left(\tau-\tau^{\prime}\right)=\left[\widehat{b}\left(\tau\right),\widehat{b}^{\dagger}\left(\tau^{\prime}\right)\right]=\intop_{0}^{+\infty}d\omega\left|c\left(\omega\right)\right|^{2}\exp\left(-i\omega\left(\tau-\tau^{\prime}\right)\right).\label{eq:bath_memory_function}
\end{equation}

Suppose that initially the open system is in a pure state $\left|\psi\left(0\right)\right\rangle _{\textrm{s}}$,
and the environment is initially in its ground state $\left|0\right\rangle _{\textrm{b}}$.
After time $t$, the joint system-bath state $\left|\Psi\left(t\right)\right\rangle $
will evolve according to the time-ordered exponential
\begin{equation}
\left|\Psi\left(t\right)\right\rangle =\mathcal{T}\exp\left[-i\intop_{0}^{t}d\tau\widehat{H}\left(\tau\right)\right]\left|0\right\rangle _{\textrm{b}}\otimes\left|\psi\left(0\right)\right\rangle _{\textrm{s}}.\label{eq:time-ordered-exponent-total-evolution}
\end{equation}
Here and below the subscripts ``s'' and ``b'' denote the states
from the Hilbert space of the system and of the bath correspondingly.
The absence of subscript means that the ket or bra vector belongs
to the total (joint) Hilbert space. 

\subsection{\label{subsec:Reduced-density-matrix}Reduced density matrix as a
statistical mixture of the bath measurement outcomes}

At the time moment $t$, the state of OQS is described by a reduced
density matrix
\begin{equation}
\widehat{\rho}_{\textrm{s}}\left(t\right)=\textrm{Tr}_{\textrm{b}}\left|\Psi\left(t\right)\right\rangle \left\langle \Psi\left(t\right)\right|,\label{eq:reduced_density}
\end{equation}
where $\textrm{Tr}_{\textrm{b}}$ is a partial trace over the bath
degrees of freedom. According to our picture, the trace operation
$\textrm{Tr}_{\textrm{b}}$ is acting on the observable quanta of
the bath. Since there can be large number of such quanta, so that
the resulting quantum state is complex, we want to calculate $\textrm{Tr}_{\textrm{b}}$
by a Monte Carlo sampling. For this purpose, we employ the (unnormalized)
coherent states of the bath 
\begin{equation}
\left\langle z\right|_{\textrm{b}}=\left\langle 0\right|_{\textrm{b}}\exp\left[\int d\omega z\left(\omega\right)\widehat{a}\left(\omega\right)\right],
\end{equation}
which depend on complex functions of frequency $z\left(\omega\right)$,
and which possess the resulution of unity property \footnote{here $\int D\left[z\right]$ is understood as the limit $\int D\left[z\right]=\lim_{M\to\infty}\frac{1}{\pi^{M}}\int dz_{x}\left(\omega_{1}\right)dz_{y}\left(\omega_{1}\right)\ldots\int dz_{x}\left(\omega_{M}\right)dz_{y}\left(\omega_{M}\right)$
for a discretization of the frequency axis $\omega_{1},\ldots\omega_{M}$,
and $z_{x}\left(\omega\right),z_{y}\left(\omega\right)$ are understood
as real and imaginary parts correspondingly of the complex number
$z\left(\omega\right)$.}
\begin{equation}
\widehat{1}_{\textrm{b}}=\int D\left[z\right]\left|z\right\rangle _{\textrm{b}}\left\langle z\right|_{\textrm{b}}\exp\left[-\int d\omega\left|z\left(\omega\right)\right|^{2}\right],\label{eq:resolution_of_unity-1}
\end{equation}
where $\widehat{1}_{\textrm{b}}$ is the identity operator in the
Hilbert space of the bath. We substitute the resolution of unity (\ref{eq:resolution_of_unity-1})
into the reduced density matrix expression (\ref{eq:reduced_density}):
\begin{equation}
\widehat{\rho}_{\textrm{s}}\left(t\right)=\int D\left[z\right]\left\langle z\right|_{\textrm{b}}\left(\left|\Psi\left(t\right)\right\rangle \left\langle \Psi\left(t\right)\right|\right)\left|z\right\rangle _{\textrm{b}}\exp\left[-\int d\omega\left|z\left(\omega\right)\right|^{2}\right].\label{eq:trace_through_the_coherent_states}
\end{equation}
Here, the \textit{partially} projected wavefunctions are introduced
\begin{equation}
\left|\psi\left(z,t\right)\right\rangle _{\textrm{s}}=\left\langle z\right|_{\textrm{b}}\left|\Psi\left(t\right)\right\rangle \coloneqq\sum_{q}\left|q\right\rangle _{\textrm{s}}\left(\left\langle z\right|_{\textrm{b}}\otimes\left\langle q\right|_{\textrm{s}}\right)\left|\Psi\left(t\right)\right\rangle ,\label{eq:projected_state}
\end{equation}
where $\left|q\right\rangle _{\textrm{s}}$ is arbitrary basis of
OQS states. The state $\left|\psi\left(z,t\right)\right\rangle _{\textrm{s}}$
is a pure state in the Hilbert space of OQS. It no longer depends
on the bath degrees of freedom. Equation (\ref{eq:trace_through_the_coherent_states})
can be interpreted in the following way. At the time moment $t$ we
measure the bath state, and find it a coherent state corresponding
to the signal $z\left(\omega\right)$. The probability to observe
a particular outcome $z\left(\omega\right)$ is provided by the real
positive Husimi function 
\begin{equation}
Q\left(z,t\right)=\textrm{Tr}_{\textrm{s}}\left\langle z\right|_{\textrm{b}}\left(\left|\Psi\left(t\right)\right\rangle \left\langle \Psi\left(t\right)\right|\right)\left|z\right\rangle _{\textrm{b}}\exp\left[-\int d\omega\left|z\left(\omega\right)\right|^{2}\right].\label{eq:Husimi_definition}
\end{equation}
In terms of Husimi function, the reduced density matrix (\ref{eq:trace_through_the_coherent_states})
assumes the form
\begin{equation}
\widehat{\rho}_{\textrm{s}}\left(t\right)=\int D\left[z\right]\left\langle z\right|_{\textrm{b}}\frac{\left\langle z\right|_{\textrm{b}}\left(\left|\Psi\left(t\right)\right\rangle \left\langle \Psi\left(t\right)\right|\right)\left|z\right\rangle _{\textrm{b}}}{\left\Vert \left\langle z\right|_{\textrm{b}}\left|\Psi\left(t\right)\right\rangle \right\Vert ^{2}}Q\left(z,t\right),\label{eq:Husimi_average}
\end{equation}
where the projection 
\begin{equation}
\widehat{\rho}_{\textrm{s}}\left(\left.t\right|z\right)=\frac{\left\langle z\right|_{\textrm{b}}\left(\left|\Psi\left(t\right)\right\rangle \left\langle \Psi\left(t\right)\right|\right)\left|z\right\rangle _{\textrm{b}}}{\left\Vert \left\langle z\right|_{\textrm{b}}\left|\Psi\left(t\right)\right\rangle \right\Vert ^{2}}\label{eq:conditional_density_matrix}
\end{equation}
can be interpreted as a pure density matrix of OQS conditioned on
a particular outcome $z\left(\omega\right)$. 

Now, provided we are able to evaluate $\widehat{\rho}_{\textrm{s}}\left(\left.t\right|z\right)$
for given $t$ and $z\left(\omega\right)$, we can devise a Monte
Carlo procedure by sampling stochastically the realizations $z^{\left(i\right)}$
of $z$ from $Q\left(z,t\right)$, and performing the average
\begin{equation}
\widehat{\rho}_{\textrm{s}}\left(t\right)=\frac{1}{M}\sum_{i=1}^{M}\widehat{\rho}_{\textrm{s}}\left(\left.t\right|z^{\left(i\right)}\right),
\end{equation}
where $M$ is the number of noise samples $z^{\left(i\right)}$. Therefore,
the following sections are devoted to the derivation of equations
of motion for $\widehat{\rho}_{\textrm{s}}\left(\left.t\right|z\right)$
and $Q\left(z,t\right)$. 

The exposition of this section was completely in line with the NMQSD
approach \citep{Diosi1998}. However in the following section, a difference
will arise. 

\subsection{Dressed state of open quantum system\label{subsec:Dressed-non-Markovian-quantum-1}}

In this section we derive the equation of motions for the conditional
OQS pure state $\widehat{\rho}_{\textrm{s}}\left(\left.t\right|z\right)$,
or, which is equivalent, for the projection $\left|\psi\left(z,t\right)\right\rangle _{\textrm{s}}$.
Since the observable state of the bath is taken into account by a
stochastic sampling from its Husimi function $Q\left(z,t\right)$,
we expect that $\left|\psi\left(z,t\right)\right\rangle _{\textrm{s}}$
will contain only unobservable, or virtual, quantum bath contributions.
This will become evident in the course of exposition.

In order to derive the equation of motion for $\left|\psi\left(z,t\right)\right\rangle _{\textrm{s}}$,
we insert the ordered operator exponent (\ref{eq:time-ordered-exponent-total-evolution})
into (\ref{eq:projected_state}):
\begin{multline}
\left|\psi\left(z,t\right)\right\rangle _{\textrm{s}}=\left\langle z\right|_{\textrm{b}}\mathcal{T}\exp\left[-i\intop_{0}^{t}d\tau\widehat{H}\left(\tau\right)\right]\left|0\right\rangle _{\textrm{b}}\otimes\left|\psi\left(0\right)\right\rangle _{\textrm{s}}\\
=\left\langle 0\right|_{\textrm{b}}\exp\left[\int d\omega z\left(\omega\right)\widehat{a}\left(\omega\right)\right]\mathcal{T}\exp\left[-i\intop_{0}^{t}d\tau\widehat{H}\left(\tau\right)\right]\left|0\right\rangle _{\textrm{b}}\otimes\left|\psi\left(0\right)\right\rangle _{\textrm{s}}.
\end{multline}
In the last line of these equations, we have a product of the two
operator exponents. Then, we transform this expression by commuting
the left exponent through the right one by employing the following
commutation relations:
\begin{equation}
\exp\left[\int d\omega z\left(\omega\right)\widehat{a}\left(\omega\right)\right]\widehat{b}\left(\tau\right)=\widehat{b}\left(\tau\right)\exp\left[\int d\omega z\left(\omega\right)\widehat{a}\left(\omega\right)\right]
\end{equation}
and
\begin{equation}
\exp\left[\int d\omega z\left(\omega\right)\widehat{a}\left(\omega\right)\right]\widehat{b}^{\dagger}\left(\tau\right)=\left[\widehat{b}^{\dagger}\left(\tau\right)+\xi\left(\tau\right)\right]\exp\left[\int d\omega z\left(\omega\right)\widehat{a}\left(\omega\right)\right],
\end{equation}
where the classical time-dependent complex noise term
\begin{equation}
\xi\left(\tau\right)=\intop_{0}^{+\infty}d\omega c^{*}\left(\omega\right)z\left(\omega\right)e^{i\omega\tau}
\end{equation}
is a consequence of the canonical commutation relations (\ref{eq:canonical_commutation_relations}).
The resulting expression for the projected state is 
\begin{equation}
\left|\psi\left(z,t\right)\right\rangle _{\textrm{s}}=\left\langle 0\right|_{\textrm{b}}\mathcal{T}\exp\left[-i\intop_{0}^{t}d\tau\widehat{H}_{\textrm{dress}}\left(z,\tau\right)\right]\left|0\right\rangle _{\textrm{b}}\otimes\left|\psi\left(0\right)\right\rangle _{\textrm{s}},\label{eq:projected_state_as_vacuum_vacuum_evolution}
\end{equation}
where the non-Hermitian Hamiltonian is
\begin{equation}
\widehat{H}_{\textrm{dress}}\left(z,\tau\right)=\widehat{H}_{\textrm{s}}+\hat{s}\left(\xi\left(\tau\right)+\widehat{b}^{\dagger}\left(\tau\right)\right)+\hat{s}^{\dagger}\widehat{b}\left(\tau\right).\label{eq:dressed_Hamiltonian}
\end{equation}

Observe that in Eq. (\ref{eq:projected_state_as_vacuum_vacuum_evolution})
we begin the evolution from the vacuum of bath, and at the time of
measurement $t$ the state is again projected to the vacuum. This
means that all the quanta which were emitted via $\widehat{b}^{\dagger}\left(\tau\right)$
at earlier times $\tau$, will be ultimately reabsorbed by $\widehat{b}\left(\tau^{\prime}\right)$
at later times $\tau^{\prime}$. In other words, they comprise the
\textit{virtual} quantum field of the bath, which is never observed
directly. At the same time, all the quanta which survive up to the
measurement time (the \textit{observable} quantum field), will be
projected onto the coherent state, thus collapsing to the classical
field $\xi\left(\tau\right)$.

We obtain the following physical picture. Given a fixed outcome $z$
for the measurement of the observable quantum field, the joint state
$\left|\Psi_{\textrm{dress}}\left(z;t\right)\right\rangle $ of OQS
and of the virtual quantum field evolves according to the Schrodinger
equation 
\begin{equation}
\left|\Psi_{\textrm{dress}}\left(z,t\right)\right\rangle =-i\widehat{H}_{\textrm{dress}}\left(z,t\right)\left|\Psi_{\textrm{dress}}\left(z,t\right)\right\rangle ,\label{eq:dressed_non_markovian_quantum_state_diffusion}
\end{equation}
with the initial condition 
\begin{equation}
\left|\Psi_{\textrm{dress}}\left(z,0\right)\right\rangle =\left|0\right\rangle _{\textrm{b}}\otimes\left|\psi\left(0\right)\right\rangle _{\textrm{s}}.\label{eq:dnmqsd_initial_condition}
\end{equation}
Now, let us again return to the statistical interpretation of the
reduced density matrix (\ref{eq:Husimi_average}). In order to compute
the reduced density matrix of the OQS, we discard all the virtual
quanta, by\textit{ partially} projecting to the bath vacuum, and then
average over all the possible measurement outcomes, with the probability
distribution $Q\left(z,t\right)=\left\Vert \left|\Psi_{\textrm{dress}}\left(z,0\right)\right\rangle \right\Vert ^{2}$:
\begin{equation}
\widehat{\rho}_{\textrm{s}}\left(t\right)=\overline{\left\{ \frac{\left\langle 0\right|_{\textrm{b}}\left|\Psi_{\textrm{dress}}\left(z,t\right)\right\rangle \left\langle \Psi_{\textrm{dress}}\left(z,t\right)\right|\left|0\right\rangle _{\textrm{b}}}{\left\Vert \left|\Psi_{\textrm{dress}}\left(z,0\right)\right\rangle \right\Vert ^{2}}\right\} }_{z}.\label{eq:density_matrix_averaged_over_outcomes}
\end{equation}
We call the joint state $\left|\Psi_{\textrm{dress}}\left(z;t\right)\right\rangle $
the (unnormalized) \textit{dressed state} of the open system. In order
to acknowledge and distunguish from the related developments in literature
\citep{Diosi1998}, we call the equation (\ref{eq:dressed_non_markovian_quantum_state_diffusion})
the \textit{dressed} NMQSD. 

The phenomenon of ``dressing'', when a small system is interacting
with the surrounding medium, is ubiquitous in physics. The prominent
examples are: the dressing of the ``bare'' electron in QED by a
cloud of virtual photons \citep{Milonni1993}; electron in a solid
crystal gets dressed by a cloud of virtual phonons, thus forming a
polaron quasiparticle \citep{Holstein1959}. What is important to
observe is that in all these cases the dressing and the ``bare''
system form a single physical object, whose properties are changed
(renormalized) as compared to the ``bare'' case. Thus, in this work
we also take on the view that $\left|\Psi_{\textrm{dress}}\left(z;t\right)\right\rangle $
is the most natural characterization of OQS state, when the bath is
non-Markovian.

These considerations allow us to anticipate the complexity properties
of the description in terms of dressed OQS state. Indeed, if the cloud
of virtual photons has the meaning of (retarded) polarization of the
bath around OQS, it is not expected to explode, or to grow without
bounds, as the time increases. Instead, it is expected to saturate
on a certain average number of quanta, even at large times. This makes
this approach attractive for the numerical simulation algorithms:
we can hope to achieve an accurate solution of dressed NMQSD (\ref{eq:dressed_non_markovian_quantum_state_diffusion})
on wide time scales, by using a Fock basis for the virtual quanta,
which is truncated above certain occupation numbers. 

At the same time, the observable quantum fields have different complexity
behaviour: since OQS can continuously emit excitations, especially
if it is driven, or the coupling with the bath is non-RWA, their occupation
numbers can grow without bounds with time. Therefore, the observable
quantum fields are the major factor which makes the real time simulation
such a hard problem. And this is the virtue of the presented approach
that these complicated quantum fields are simulated stochastically
through the classical field $\xi\left(\tau\right)$. 

In conculsion of this section, we note that the vacuum is not the
only initial condition for the bath $\left|0\right\rangle _{\textrm{b}}$.
In particular, by adding a suitable stochastic field, a thermal initial
state of the bath can be considered \citep{Hartmann2017,Polyakov2017b,Polyakov2017a}

\subsection{\label{subsec:Master-equation-for}Master equation for Husimi function
and its stochastic interpretation}

Now, having all the means to compute the conditional state $\widehat{\rho}_{\textrm{s}}\left(\left.t\right|z\right)$,
the remaining task is devise a procedure for the stochastic sampling
from the Husimi function $Q\left(z,t\right)$. Here we will follow
a route which is analogous to that in the conventional NMQSD, but
taking into account the specific features of our definition of the
quantum trajectory.

We start the derivation by computing the time derivative of the Husimi
function:

\begin{multline}
\partial_{t}Q\left(z,t\right)=\exp\left(-\int d\omega\left|z\left(\omega\right)\right|^{2}\right)\left\langle \Psi_{\textrm{dress}}\left(z;t\right)\right|\left\{ \overleftarrow{\partial}_{t}\left|0\right\rangle _{\textrm{b}}+\left\langle 0\right|_{\textrm{b}}\overrightarrow{\partial}_{t}\right\} \left|\Psi_{\textrm{dress}}\left(z,t\right)\right\rangle \\
=i\exp\left(-\int d\omega\left|z\left(\omega\right)\right|^{2}\right)\left\langle \Psi_{\textrm{dress}}\left(z;t\right)\right|\left\{ \widehat{H}_{\textrm{dress}}\left(z,t\right)\left|0\right\rangle _{\textrm{b}}-\left\langle 0\right|_{\textrm{b}}\widehat{H}_{\textrm{dress}}\left(z,t\right)\right\} \left|\Psi_{\textrm{dress}}\left(z,t\right)\right\rangle .\label{eq:time_derivative_Husimi}
\end{multline}
Upon substituting the expression (\ref{eq:dressed_Hamiltonian}) for
$\widehat{H}_{\textrm{dress}}$, we obtain a number of terms. Then,
we employ the crucial propery of the dressed state: it has the following
functional dependence on the virtual operators and on the noise: 
\begin{equation}
\left|\Psi_{\textrm{dress}}\left(z,t\right)\right\rangle \equiv\Phi\left(z\left(\omega\right)+\widehat{b}^{\dagger}\left(\omega\right),\widehat{b}\left(\omega\right),t\right)\left|0\right\rangle _{\textrm{b}},\label{eq:functional_dependence_of_dressed_state}
\end{equation}
so that
\begin{equation}
\widehat{b}\left(\omega\right)\left|\Psi_{\textrm{dress}}\left(z,t\right)\right\rangle =\frac{\delta}{\delta z\left(\omega\right)}\left|\Psi_{\textrm{dress}}\left(z,t\right)\right\rangle .\label{eq:annihilator_as_noise_derivative}
\end{equation}
Using these properties, we arrange all the terms in (\ref{eq:time_derivative_Husimi}),
and obtain the following master equation:
\begin{equation}
\partial_{t}Q\left(z,t\right)=\int d\omega\frac{\delta}{\delta z\left(\omega\right)}\left\{ ie^{+i\omega t}c\left(\omega\right)\overline{s}\left(z\right)Q\left(z,t\right)\right\} +\int d\omega\frac{\delta}{\delta z\left(\omega\right)}\left\{ -ie^{-i\omega t}c\left(\omega\right)\overline{s}^{*}\left(z\right)Q\left(z,t\right)\right\} ,\label{eq:Husimi_master_equation}
\end{equation}
where the average of the system operator is defined as 
\begin{equation}
\overline{s}\left(z,t\right)=\textrm{Tr}_{\textrm{s}}\left\{ \widehat{s}\widehat{\rho}_{\textrm{s}}\left(\left.t\right|z\right)\right\} =\frac{\left\langle \Psi_{\textrm{dress}}\left(z,t\right)\left|0\right.\right\rangle _{\textrm{b}}\widehat{s}\left\langle 0\right|_{\textrm{b}}\left|\Psi_{\textrm{dress}}\left(z,t\right)\right\rangle }{\left\Vert \left\langle 0\right|_{\textrm{b}}\left|\Psi_{\textrm{dress}}\left(z,t\right)\right\rangle \right\Vert ^{2}}.\label{eq:conditional_system_operator_average}
\end{equation}
The interested reader can find the technical details of the derivation
in the Appendix \ref{sec:NORMALIZED-HUSIMI-QUANTUM}. Here we turn
to the interpretation of the master equation. Formally, this equation
corresponds to convection. Indeed, we have an infinite-dimensional
space of all the possible signal realizations $z\left(\omega\right)$.
We also have an initial probability distribution in this space 
\begin{equation}
Q\left(z,0\right)=\exp\left[-\int d\omega\left|z\left(\omega\right)\right|^{2}\right],\label{eq:initial_Husimi}
\end{equation}
which follows from (\ref{eq:Husimi_definition}). Finally, for every
time moment $t\geqslant0$ and for every fixed signal $z$, there
is a complex velocity field
\begin{equation}
v\left(\omega,z,t\right)=ie^{-i\omega t}c\left(\omega\right)\textrm{Tr}_{\textrm{s}}\left\{ \widehat{s}^{\dagger}\widehat{\rho}_{\textrm{s}}\left(\left.t\right|z\right)\right\} 
\end{equation}
for the component $z\left(\omega\right)$ of the signal. Indeed, this
velocity value can be evaluated explicitly by solving the Schrodinger
equation (\ref{eq:dressed_non_markovian_quantum_state_diffusion})
with the inital conditions (\ref{eq:dnmqsd_initial_condition}). Then,
the master equation (\ref{eq:Husimi_master_equation}) corresponds
to a convection with this velocity field:
\begin{equation}
\partial_{t}z\left(\omega\right)=ic\left(\omega\right)e^{-i\omega t}\overline{s}^{*}\left(z,t\right).\label{eq:convetion_equation}
\end{equation}
Solving this equation, we obtain the displacement of the signal at
a time $t$: 
\begin{equation}
z\left(\omega,t\right)=z\left(\omega,0\right)+ic\left(\omega\right)\intop_{0}^{t}e^{-i\omega\tau}\overline{s}^{*}\left(z\left(\tau\right),\tau\right)d\tau,\label{eq:evolved_noise-1}
\end{equation}
where $z\left(\tau\right)$ denotes all the components $z\left(\omega,\tau\right)$.
In terms of the open system and the bath, this result means that the
observable bath displacement is due to non-zero average value of the
system operator. 

Now we have a conceptual understanding of how to conduct a stochastic
sampling from $Q\left(z,0\right)$. First, we draw the initial conditions
$z\left(\omega,0\right)$ from the initial Gaussian distribution (\ref{eq:initial_Husimi}).
Then, we evolve them up to the time $t$ according to (\ref{eq:evolved_noise-1}).
Finally, the reduced OQS density matrix is caclulated as the average:
\begin{equation}
\widehat{\rho}_{\textrm{s}}\left(t\right)=\overline{\left\{ \widehat{\rho}_{\textrm{s}}\left(\left.t\right|z\left(\omega,t\right)\right)\right\} }_{z\left(\omega,0\right)}.
\end{equation}

\subsection{\label{subsec:Husimi-dressed-quantum}Husimi dressed quantum trajectories}

The last thing to do before we arrive at the final simulation algorithm
is to derive the explicit equation for the joint self-consistent evolution
of $z\left(\omega,t\right)$ and $\widehat{\rho}_{\textrm{s}}\left(\left.t\right|z\left(\omega,t\right)\right)$.
We have for the time derivative of $\left|\Psi_{\textrm{dress}}\left(z\left(t\right),t\right)\right\rangle $:
\begin{equation}
\frac{d}{dt}\left|\Psi_{\textrm{dress}}\left(z\left(t\right),t\right)\right\rangle =\partial_{t}\left|\Psi_{\textrm{dress}}\left(z\left(t\right),t\right)\right\rangle +\int d\omega\dot{z}\left(\omega,t\right)\frac{\delta}{\delta z\left(\omega,t\right)}\left|\Psi_{\textrm{dress}}\left(z\left(t\right),t\right)\right\rangle .
\end{equation}
The time derivative $\dot{z}\left(\omega,t\right)$ is given by the
convection equation (\ref{eq:convetion_equation}). Since the properties
(\ref{eq:functional_dependence_of_dressed_state})-(\ref{eq:annihilator_as_noise_derivative})
hold when the noise is time-dependent, we make the substitution 
\begin{equation}
\frac{\delta}{\delta z\left(\omega,t\right)}\left|\Psi_{\textrm{dress}}\left(z\left(t\right),t\right)\right\rangle =\widehat{b}\left(\omega\right)\left|\Psi_{\textrm{dress}}\left(z\left(t\right),t\right)\right\rangle ,
\end{equation}
and obtain the following closed self-consistent Schrodinger equation
\begin{equation}
\left|\Psi_{\textrm{dress}}\left(z\left(t\right),t\right)\right\rangle =-i\widehat{H}_{\textrm{dress}}^{\prime}\left(z,t\right)\left|\Psi_{\textrm{dress}}\left(z\left(t\right),t\right)\right\rangle ,\label{eq:shifted_Schrodinger_equation-1}
\end{equation}
where the Hamiltonian is
\begin{equation}
\widehat{H}_{\textrm{dress}}^{\prime}\left(z,t\right)=\widehat{H}_{\textrm{s}}+\hat{s}\left(\xi\left(t\right)+\phi^{*}\left(t\right)+\widehat{b}^{\dagger}\left(t\right)\right)+\left(\hat{s}^{\dagger}-\overline{s}^{*}\left(t\right)\right)\widehat{b}\left(t\right).\label{eq:dressed_nonlinear_nonmarkovian_quantum_state_diffusion-1}
\end{equation}
Here the system operator average $\overline{s}\left(t\right)$ is
\begin{equation}
\overline{s}\left(t\right)=\frac{\left\langle \Psi_{\textrm{dress}}\left(z\left(t\right),t\right)\left|0\right.\right\rangle _{\textrm{b}}\widehat{s}\left\langle 0\right|_{\textrm{b}}\left|\Psi_{\textrm{dress}}\left(z\left(t\right),t\right)\right\rangle }{\left\Vert \left\langle 0\right|_{\textrm{b}}\left|\Psi_{\textrm{dress}}\left(z\left(t\right),t\right)\right\rangle \right\Vert ^{2}},
\end{equation}
and the self-consistent field $\phi\left(t\right)$ is a retarded
convolution of the history of values of $\overline{s}\left(t\right)$
\begin{equation}
\phi\left(t\right)=-i\intop_{0}^{t}d\tau M\left(t-\tau\right)\overline{s}\left(\tau\right).
\end{equation}
The equations (\ref{eq:shifted_Schrodinger_equation-1})-(\ref{eq:dressed_nonlinear_nonmarkovian_quantum_state_diffusion-1})
may be called the dressed \textit{non-linear} NMQSD, in analogy with
the corresponding developments in the literature \citep{Diosi1998}.
However, a better name would be the Husimi dressed quantum trajectory,
since these trajectories provide a stochastic interpretation of the
master equation for the Husimi phase-space distribution. 

\section{NUMERICAL ALGORITHM AND EXAMPLE CALCULATION}

\subsection{\label{subsec:Summary-of-the}Summary of the resulting numerical
scheme}

For a given open quantum system (\ref{eq:total_hamiltonian_schrodinger_picture}),
(\ref{eq:free_bath_hamiltonian}), (\ref{eq:coupling_to_the_bath}),
the bath frequency range $\left[0,\omega_{\textrm{max}}\right]$ is
chosen, and is discretized in a certain way $\omega_{1}$,..., $\omega_{N}$.
The bath mode operators are approximated as
\begin{equation}
\widehat{a}_{i}\approx\sqrt{\Delta\omega}\widehat{a}\left(\omega_{i}\right),
\end{equation}
with the resulting commutation relations $\left[\widehat{a}_{i},\widehat{a}_{k}^{\dagger}\right]=\delta_{ik}$,
so that
\begin{equation}
\widehat{b}\approx\sum_{i=1}^{N}\sqrt{\Delta\omega}c\left(\omega_{i}\right)\widehat{a}_{i}\,\,\,\textrm{and}\,\,\,\widehat{H}_{\textrm{b}}\approx\sum_{i=1}^{N}\omega_{i}\widehat{a}_{i}^{\dagger}\widehat{a}_{i}.
\end{equation}
In a numerical simulation, we draw $M$ samples of the discretized
signal $\boldsymbol{z}^{\left(1\right)},\ldots\boldsymbol{z}^{\left(M\right)}$,
where now $j$th sample $\boldsymbol{z}^{\left(j\right)}$ is the
complex vector of $N$ components:
\begin{equation}
\boldsymbol{z}^{\left(j\right)}=\left[\begin{array}{c}
z_{1}^{\left(j\right)}\\
\vdots\\
z_{N}^{\left(j\right)}
\end{array}\right],
\end{equation}
and $z_{i}^{\left(j\right)}$ corresponds to the frequency $\omega_{i}$.
The distribution of samples is complex Gaussian, with the statistics
\begin{equation}
\overline{z_{i}}=\overline{z_{i}^{*}}=\overline{z_{i}^{2}}=\overline{z_{i}^{2*}}=0,\,\,\,\overline{z_{i}z_{k}^{*}}=\delta_{ik}.
\end{equation}
For each sample $\boldsymbol{z}^{\left(j\right)}$, we solve for the
Husimi quantum trajectory, 
\begin{equation}
\left|\Psi_{Q}^{\left(j\right)}\left(t\right)\right\rangle =-i\widehat{H}_{Q}\left(\boldsymbol{z}^{\left(j\right)},t\right)\left|\Psi_{Q}^{\left(j\right)}\left(t\right)\right\rangle ,
\end{equation}
where $\widehat{H}_{Q}$ is obtained from discretization of Eq. (\ref{eq:dressed_nonlinear_nonmarkovian_quantum_state_diffusion-1})
in the Schrodinger picture:
\begin{equation}
\widehat{H}_{Q}\left(\boldsymbol{z}^{\left(j\right)},t\right)=\widehat{H}_{\textrm{s}}+\hat{s}\left(\sqrt{\Delta\omega}\sum_{i=1}^{N}c^{*}\left(\omega_{i}\right)z_{i}^{\left(j\right)}e^{i\omega t}+\phi^{*}\left(t\right)+\widehat{b}^{\dagger}\right)+\left(\hat{s}^{\dagger}-\overline{s}^{*}\left(t\right)\right)\widehat{b}+\widehat{H}_{\textrm{b}}.\label{eq:numerical_Husimi_quantum_trajectory}
\end{equation}
Here the Hamiltonian $\widehat{H}_{Q}$ depends self-consistently
on the average of the system operator
\begin{equation}
\overline{s}\left(t\right)=\frac{\left\langle \Psi_{Q}^{\left(j\right)}\left(t\right)\left|0\right.\right\rangle _{\textrm{b}}\widehat{s}\left\langle 0\right|_{\textrm{b}}\left|\Psi_{Q}^{\left(j\right)}\left(t\right)\right\rangle }{\left\Vert \left\langle 0\right|_{\textrm{b}}\left|\Psi_{Q}^{\left(j\right)}\left(t\right)\right\rangle \right\Vert ^{2}}
\end{equation}
and on the retarded field $\phi\left(t\right)$ 
\begin{equation}
\phi\left(t\right)=-i\intop_{0}^{t}d\tau M\left(t-\tau\right)\overline{s}\left(\tau\right).
\end{equation}
The initial condition for the trajectory $\left|\Psi_{Q}^{\left(j\right)}\left(t\right)\right\rangle $
is
\begin{equation}
\left|\Psi_{Q}^{\left(j\right)}\left(0\right)\right\rangle =\left|0\right\rangle _{\textrm{b}}\otimes\left|\psi\left(0\right)\right\rangle _{\textrm{s}}.
\end{equation}
The equation (\ref{eq:numerical_Husimi_quantum_trajectory}) is solved
in a finite Hilbert space of virtual quanta (truncated in maximal
total occupation $n$). The result of such a simulation, the reduced
density matrix of OQS, is calculated by averaging over the signal
samples:
\begin{equation}
\widehat{\rho}_{\textrm{s}}\left(t\right)=\frac{1}{M}\sum_{j=1}^{M}\frac{\left\langle 0\right|_{\textrm{b}}\left|\Psi_{Q}^{\left(j\right)}\left(t\right)\right\rangle \left\langle \Psi_{Q}^{\left(j\right)}\left(t\right)\right|\left|0\right\rangle _{\textrm{b}}}{\left\Vert \left|\Psi_{Q}^{\left(j\right)}\left(0\right)\right\rangle \right\Vert ^{2}}.
\end{equation}

\subsection{\label{subsec:Example-calculation}Example calculation\label{subsec:Numerical-algorithm-and}}

\begin{figure*}
\includegraphics{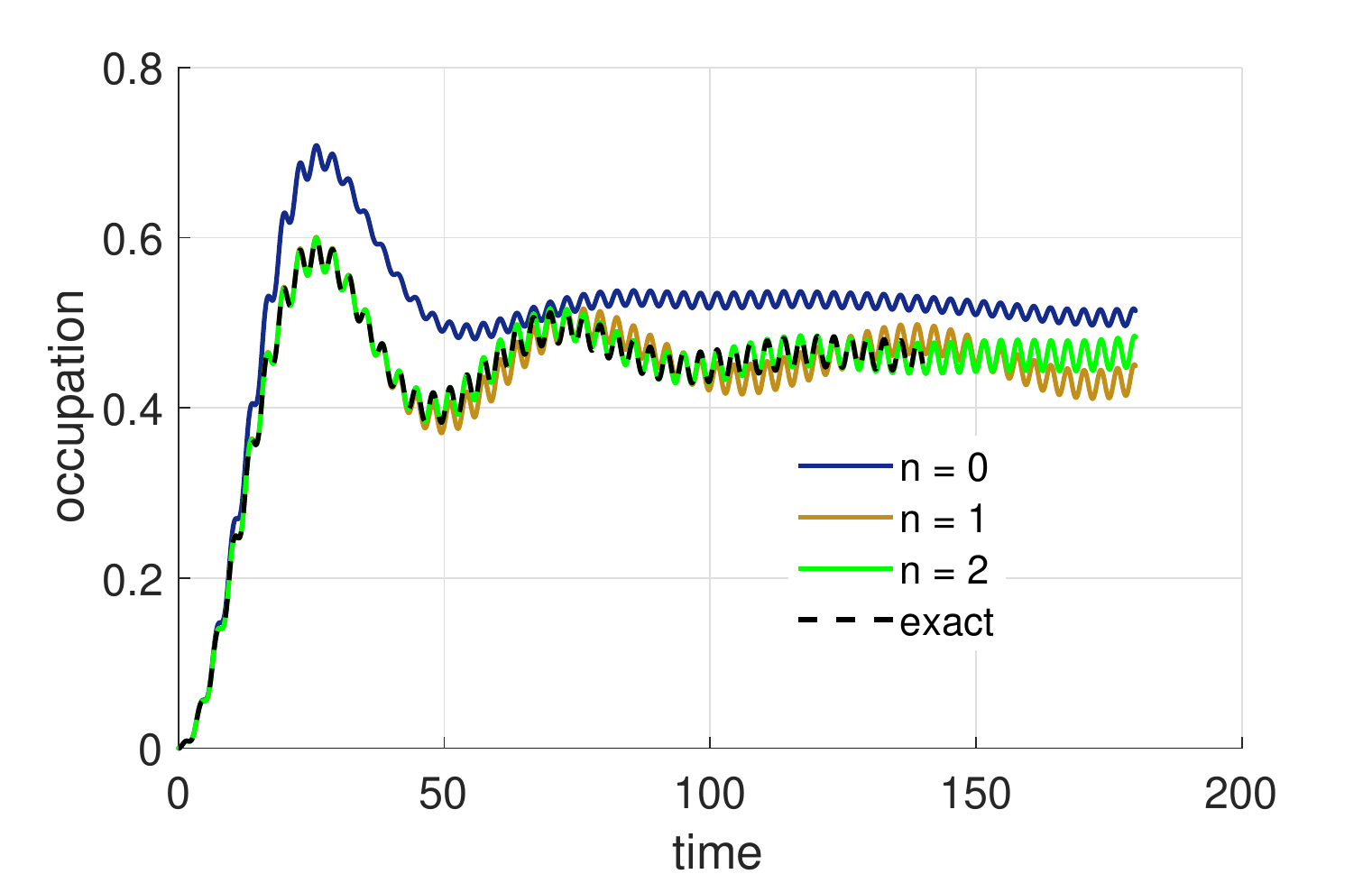}

\caption{\label{fig:figure_convergence_band_centered-1} The spin energy level
is at the center of bath's energy band. The average occupation of
the qubit is computed by exact diagonalization (in the truncated full
Fock space), which required us to take into account 8 bath excitations
to achieve the convergence up to the time $t=180$ (dahsed line).
At the same time, the results for the dressed quantum trajectory method
at $n=0$ (blue), $n=1$ (brown), and $n=2$ (green), show that we
reach convergence up to the stationary regime with only 2 virtual
excitations.}
\end{figure*}

We test the proposed approach on the spin-boson model,
\begin{equation}
\widehat{H}_{\textrm{s}}=\varepsilon\widehat{\sigma}_{+}\widehat{\sigma}_{-}+\widehat{\sigma}_{+}f\left(t\right)+\widehat{\sigma}_{-}f^{*}\left(t\right),\label{eq:driven_spin_boson}
\end{equation}
with the driving field 
\begin{equation}
f\left(t\right)=0.1\cos t.\label{eq:driving}
\end{equation}
 The spin is coupled to the bath through the spin lowering operator
\begin{equation}
\widehat{s}=\widehat{\sigma}_{-}.
\end{equation}
The bath is represented by a semiinfinite chain of bose sites with
the on-site energy $\varepsilon_{0}$ and hopping between the sites
$h$. The frequency spectrum is represented by a band $\left[\varepsilon_{0}-2h,\varepsilon_{0}+2h\right]$,
which is discretized as 
\begin{equation}
\omega_{i}=\varepsilon_{0}+2h\cos\left(\frac{i\pi}{N+1}\right),\,\,\,c\left(\omega_{i}\right)=h\sqrt{\frac{2}{\pi}}\sin\left(\frac{i\pi}{N+1}\right)\label{eq:spin_boson_coupling}
\end{equation}
for $i=1\ldots N$. For calculations, we use the following values
of parameters of the bath: $\varepsilon_{0}=1$, $h=0.05,N=20$. In
Fig. \ref{fig:figure_convergence_band_centered-1} we present the
convergence of stochastic simulation with $n=0$ (only spin Hilbert
space, no virtual quanta), $n=1$ (spin Hilbert space and one virtual
bath quantum), and $n=2$ (spin Hilbert space and two virtual bath
quanta), towards the ED solution (the full Schrodinger equation in
the truncated Fock space). From the presented results we see that
the convergence on the whole time interval is achieved with only two
virtual quanta, whereas ED required to include the states with 8 excitations
of the bath. This result confirms our idea that the virtual cloud
saturates on long times, so that this approach is indeed promising
for the development of efficient long-time simulation algorithms.

\section{\label{sec:THE-PROBLEM-OF}THE PROBLEM OF MEMORY TAILS}

\begin{figure}
\includegraphics[scale=0.55]{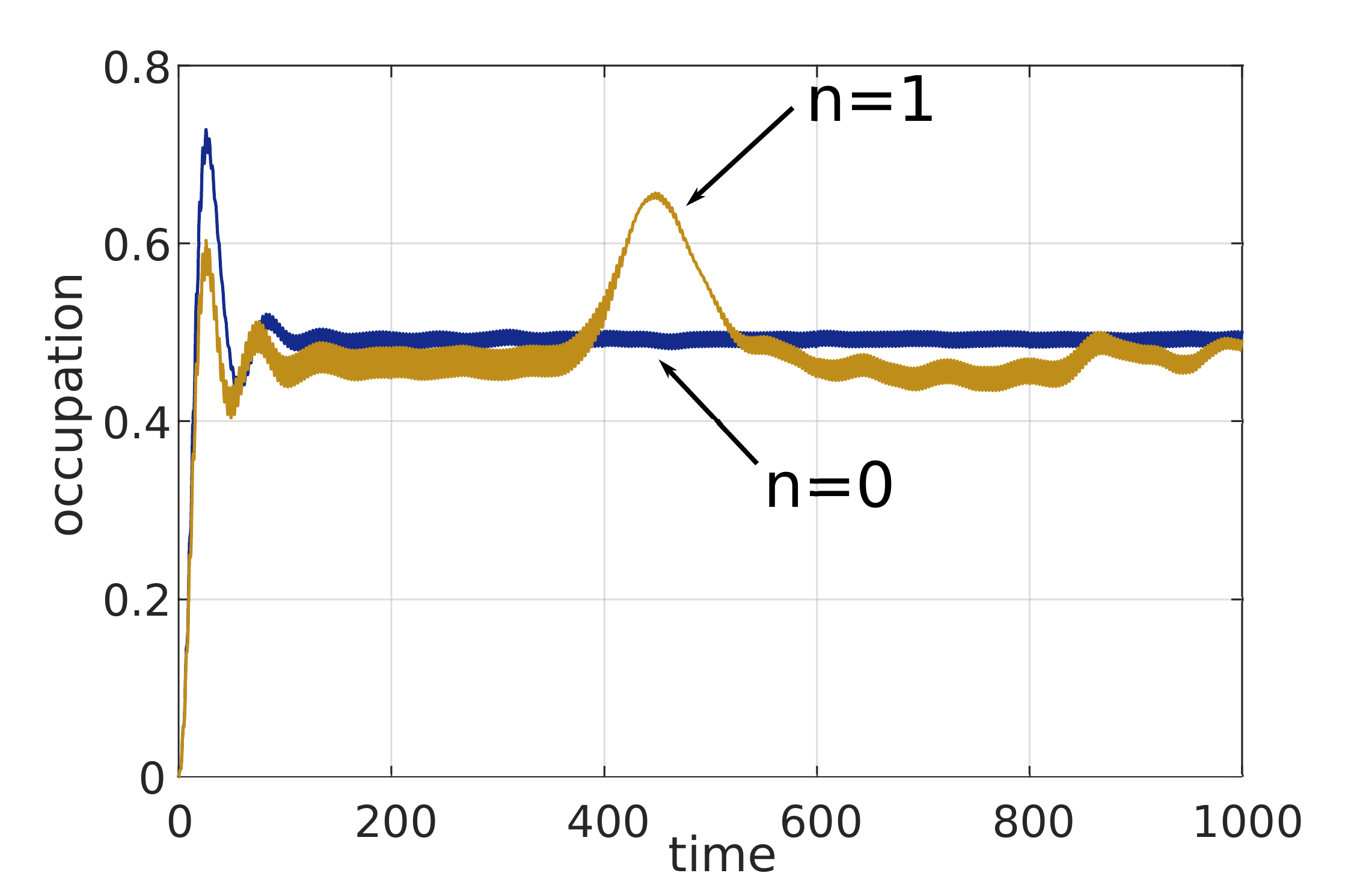}

\caption{\label{fig:long_time}The same system as in previous figure, but simulated
on much longer times. The cases of zero virtual quanta ($n=0$) and
one virtual quantum ($n=1$) are presented. It is seen that for one
virtual quantum, when a trunacted representation of the bath wavefunction
appears, a revival is observed at $t\approx400$. This is the manifestation
of the generic problem of long-range memory tails.}
\end{figure}

Let us simulate the system described in a previous section for a longer
period of time. In Fig. \ref{fig:long_time} we present the results
for zero $\left(n=0\right)$ and one $\left(n=1\right)$ virtual quantum. 

We see that for zero virtual photons the simulation is stable and
reaches the stationary non-equilibrium state. At the same time, for
one virtual quantum we observe the spurious revival around $t=400$.
Its emergence is not related to the truncation in the number of virtual
quanta $n$. Instead, for any number of virtual quanta, due to the
discretization of the bath spectrum on a finite frequency range, the
tails of the memory function $M\left(\tau\right)$ get corrupted.
The finer is the discretization, the more retarded part of the tail
is corrupted, hence the revival is more delayed. This is a general
problem of all the simulation methods. The conventional ED simulations
truncate the number of the bath modes which leads to the occurence
of the spurious reflected signal from the boundary. Other methods,
which directly deal with the memory functions, such as path integral-based
methods (QUAPI and beyond) and HOPS, directly truncate the memory
tails, which also leads to corruption of calculated observables after
a certain simulation time. 

Next paper in this series \citep{Polyakov2018b} is devoted to the
problem of memory tails. We show there that it is possible to devise
a special soft coarsegraining of the virtual quanta wavefunction,
such that revivals disappear.

\section{\label{sec:CONCULSION}CONCULSION }

In this work we propose a novel formulation for the dynamics of open
quantum systems in a non-Markovian environment. In conventional approaches,
the quantum field of the bath can develop large occupations of quanta,
with correlations of high dimensions. This presents a formiddable
obstacle both to the description of non-Markovian physics and to the
development of long-time simulation methods. Here we demonstrate that
the full quantum field can be divided into the two components, of
different physical nature, and each with its own favourable properties.
The virtual component is an intrinsically quantum object, but tends
to be asymptotically bounded at large times, so that it can be efficiently
simulated within a truncated basis. The second component of the quantum
field, the observable part, can grow without bounds with time, but
the hierarchy of its correlations have classical stochastic structure,
so that its dynamics can be efficiently simulated by Monte Carlo methods.

We believe that the proposed formulation provides us with a natural
solution to the problem of decomposition of the bath $B$ into the
entangled memory part $M$ (the virtual quanta) and the detector part
$B$ (the observable quanta). This may be useful for the analysis
of various non-Markovian phenomena such as the information and energy
backflow; for the discussion of various information and entanglement
measures in the presence non-Markovianity. 

Finally, this approach sheds light on the physical foundation of such
methods as the NMQSD and HOPS, ``explaining'' their relative success.
This may pave the way for novel, more advances, simulation techniques.
In this respect one of the favourable features of the presented formulation
is that its main concepts- the dressed state and its vacuum projection
- are simple and well-defined quantum-mechanical objects,  amenable
to analysiz, combination, or approximation by any of the numerous
conventional quantum-mechanical methods.
\begin{acknowledgments}
The study was founded by the RSF, grant 16-42-01057.
\end{acknowledgments}

\section{\label{sec:NORMALIZED-HUSIMI-QUANTUM}DERIVATION OF THE HUSIMI MASTER
EQUATION}

In the time derivative (\ref{eq:time_derivative_Husimi}) the following
terms will be zero: 
\begin{equation}
\ldots\times\left\langle 0\right|_{\textrm{b}}\widehat{b}^{\dagger}\left|\Psi_{\textrm{dress}}\left(z,t\right)\right\rangle =0\label{eq:zero_term_1}
\end{equation}
and 
\begin{equation}
\left\langle \Psi_{\textrm{dress}}\left(z,t\right)\right|\widehat{b}\left|0\right\rangle _{\textrm{b}}\times\ldots=0.\label{eq:zero_term_2}
\end{equation}
There will be also terms of the form 
\begin{equation}
\ldots\times\left\langle 0\right|_{\textrm{b}}\widehat{b}\left|\Psi_{\textrm{dress}}\left(z,t\right)\right\rangle \,\,\,\textrm{and}\,\,\,\left\langle \Psi_{\textrm{dress}}\left(z,t\right)\right|\widehat{b}^{\dagger}\left|0\right\rangle _{\textrm{b}}\times\ldots,
\end{equation}
which are non-zero, and we calculate them using the property (\ref{eq:annihilator_as_noise_derivative}).
Another property we will use is that the quantum trajectory is analytic
function of the noise: 
\begin{equation}
\frac{\delta}{\delta z^{*}\left(\omega\right)}\left|\Psi_{\textrm{dress}}\left(z,t\right)\right\rangle =0\,\,\,\textrm{and}\,\,\,\frac{\delta}{\delta z\left(\omega\right)}\left\langle \Psi_{\textrm{dress}}\left(z,t\right)\right|=0.
\end{equation}
Taking into account all these facts, we obtain for the time derivative
of $Q\left(z;t\right)$:
\begin{equation}
\partial_{t}Q\left(z,t\right)=\exp\left(-\int d\omega\left|z\left(\omega\right)\right|^{2}\right)\left\{ -iz\overline{s}_{\textrm{u}}+i\partial_{z^{*}\left(t\right)}\overline{s}_{\textrm{u}}+iz^{*}\overline{s}_{\textrm{u}}^{*}-i\partial_{z\left(t\right)}\overline{s}_{\textrm{u}}^{*}\right\} ,\label{eq:Husimi_master_incomplete}
\end{equation}
where the unnormalized average of $\widehat{s}$ was introduced 
\begin{equation}
\overline{s}_{\textrm{u}}=\left\langle \Psi_{\textrm{dress}}\left(z,t\right)\right|\left|0\right\rangle _{\textrm{b}}\widehat{s}\left\langle 0\right|_{\textrm{b}}\left|\Psi_{\textrm{dress}}\left(z,t\right)\right\rangle ,
\end{equation}
and the formal time derivative with respect to ``instantaneous''
noise value was defined (just for the brevity of notation)
\begin{equation}
\partial_{z\left(t\right)}\eqqcolon\intop_{-\infty}^{+\infty}d\omega c\left(\omega\right)\exp\left(-i\omega t\right)\frac{\delta}{\delta z\left(\omega\right)}.
\end{equation}
To proceed further from Eq. (\ref{eq:Husimi_master_incomplete}),
we employ the properties
\begin{equation}
\partial_{z\left(t\right)}\int d\omega\left|z\left(\omega\right)\right|^{2}=z^{*}\left(t\right)
\end{equation}
and
\begin{equation}
\partial_{z\left(t\right)}\exp\left(-\int d\omega\left|z\left(\omega\right)\right|^{2}\right)=\exp\left(-\int d\omega\left|z\left(\omega\right)\right|^{2}\right)\left\{ -z^{*}\left(t\right)+\partial_{z\left(t\right)}\right\} .
\end{equation}
Finally, we arrive at the master equation (\ref{eq:Husimi_master_equation}).

\bibliographystyle{apsrev4-1}
\bibliography{references}

\begin{thebibliography}{66}%
\makeatletter
\providecommand \@ifxundefined [1]{%
 \@ifx{#1\undefined}
}%
\providecommand \@ifnum [1]{%
 \ifnum #1\expandafter \@firstoftwo
 \else \expandafter \@secondoftwo
 \fi
}%
\providecommand \@ifx [1]{%
 \ifx #1\expandafter \@firstoftwo
 \else \expandafter \@secondoftwo
 \fi
}%
\providecommand \natexlab [1]{#1}%
\providecommand \enquote  [1]{``#1''}%
\providecommand \bibnamefont  [1]{#1}%
\providecommand \bibfnamefont [1]{#1}%
\providecommand \citenamefont [1]{#1}%
\providecommand \href@noop [0]{\@secondoftwo}%
\providecommand \href [0]{\begingroup \@sanitize@url \@href}%
\providecommand \@href[1]{\@@startlink{#1}\@@href}%
\providecommand \@@href[1]{\endgroup#1\@@endlink}%
\providecommand \@sanitize@url [0]{\catcode `\\12\catcode `\$12\catcode
  `\&12\catcode `\#12\catcode `\^12\catcode `\_12\catcode `\%12\relax}%
\providecommand \@@startlink[1]{}%
\providecommand \@@endlink[0]{}%
\providecommand \url  [0]{\begingroup\@sanitize@url \@url }%
\providecommand \@url [1]{\endgroup\@href {#1}{\urlprefix }}%
\providecommand \urlprefix  [0]{URL }%
\providecommand \Eprint [0]{\href }%
\providecommand \doibase [0]{http://dx.doi.org/}%
\providecommand \selectlanguage [0]{\@gobble}%
\providecommand \bibinfo  [0]{\@secondoftwo}%
\providecommand \bibfield  [0]{\@secondoftwo}%
\providecommand \translation [1]{[#1]}%
\providecommand \BibitemOpen [0]{}%
\providecommand \bibitemStop [0]{}%
\providecommand \bibitemNoStop [0]{.\EOS\space}%
\providecommand \EOS [0]{\spacefactor3000\relax}%
\providecommand \BibitemShut  [1]{\csname bibitem#1\endcsname}%
\let\auto@bib@innerbib\@empty
%</preamble>
\bibitem [{\citenamefont {Zurek}(2003)}]{Zurek2003}%
  \BibitemOpen
  \bibfield  {author} {\bibinfo {author} {\bibfnamefont {W.~H.}\ \bibnamefont
  {Zurek}},\ }\href@noop {} {\bibfield  {journal} {\bibinfo  {journal} {Rev.
  Mod. Phys.}\ }\textbf {\bibinfo {volume} {75}},\ \bibinfo {pages} {715}
  (\bibinfo {year} {2003})}\BibitemShut {NoStop}%
\bibitem [{\citenamefont {Schlosshauer}(2004)}]{Schlosshauer2004}%
  \BibitemOpen
  \bibfield  {author} {\bibinfo {author} {\bibfnamefont {M.}~\bibnamefont
  {Schlosshauer}},\ }\href@noop {} {\bibfield  {journal} {\bibinfo  {journal}
  {Rev. Mod. Phys.}\ }\textbf {\bibinfo {volume} {76}},\ \bibinfo {pages}
  {1267} (\bibinfo {year} {2004})}\BibitemShut {NoStop}%
\bibitem [{\citenamefont {Blume-Kohout}\ and\ \citenamefont
  {Zurek}(2008)}]{Blume-Kohout2008}%
  \BibitemOpen
  \bibfield  {author} {\bibinfo {author} {\bibfnamefont {R.}~\bibnamefont
  {Blume-Kohout}}\ and\ \bibinfo {author} {\bibfnamefont {W.~H.}\ \bibnamefont
  {Zurek}},\ }\href@noop {} {\bibfield  {journal} {\bibinfo  {journal} {Phys.
  Rev. Lett.}\ }\textbf {\bibinfo {volume} {101}},\ \bibinfo {pages} {240405}
  (\bibinfo {year} {2008})}\BibitemShut {NoStop}%
\bibitem [{\citenamefont {Riedel}\ and\ \citenamefont
  {Zurek}(2011)}]{Riedel2011}%
  \BibitemOpen
  \bibfield  {author} {\bibinfo {author} {\bibfnamefont {C.~J.}\ \bibnamefont
  {Riedel}}\ and\ \bibinfo {author} {\bibfnamefont {W.~H.}\ \bibnamefont
  {Zurek}},\ }\href@noop {} {\bibfield  {journal} {\bibinfo  {journal} {New J.
  Phys.}\ }\textbf {\bibinfo {volume} {13}},\ \bibinfo {pages} {073038}
  (\bibinfo {year} {2011})}\BibitemShut {NoStop}%
\bibitem [{\citenamefont {Korbicz}\ \emph {et~al.}(2017)\citenamefont
  {Korbicz}, \citenamefont {Aguilar}, \citenamefont {Cwiklinski},\ and\
  \citenamefont {Horodecki}}]{Korbicz2017}%
  \BibitemOpen
  \bibfield  {author} {\bibinfo {author} {\bibfnamefont {J.~K.}\ \bibnamefont
  {Korbicz}}, \bibinfo {author} {\bibfnamefont {E.~A.}\ \bibnamefont
  {Aguilar}}, \bibinfo {author} {\bibfnamefont {P.}~\bibnamefont {Cwiklinski}},
  \ and\ \bibinfo {author} {\bibfnamefont {P.}~\bibnamefont {Horodecki}},\
  }\href@noop {} {\bibfield  {journal} {\bibinfo  {journal} {Phys. Rev. A}\
  }\textbf {\bibinfo {volume} {96}},\ \bibinfo {pages} {032124} (\bibinfo
  {year} {2017})}\BibitemShut {NoStop}%
\bibitem [{\citenamefont {Mironowicz}\ \emph {et~al.}(2017)\citenamefont
  {Mironowicz}, \citenamefont {Korbicz},\ and\ \citenamefont
  {Horodecki}}]{Mironowicz2017}%
  \BibitemOpen
  \bibfield  {author} {\bibinfo {author} {\bibfnamefont {P.}~\bibnamefont
  {Mironowicz}}, \bibinfo {author} {\bibfnamefont {J.~K.}\ \bibnamefont
  {Korbicz}}, \ and\ \bibinfo {author} {\bibfnamefont {P.}~\bibnamefont
  {Horodecki}},\ }\href@noop {} {\bibfield  {journal} {\bibinfo  {journal}
  {Phys. Rev. Lett.}\ }\textbf {\bibinfo {volume} {118}},\ \bibinfo {pages}
  {150501} (\bibinfo {year} {2017})}\BibitemShut {NoStop}%
\bibitem [{\citenamefont {Knott}\ \emph {et~al.}(2018)\citenamefont {Knott},
  \citenamefont {Tufarelli}, \citenamefont {Piano},\ and\ \citenamefont
  {Adesso}}]{Knott2018}%
  \BibitemOpen
  \bibfield  {author} {\bibinfo {author} {\bibfnamefont {P.~A.}\ \bibnamefont
  {Knott}}, \bibinfo {author} {\bibfnamefont {T.}~\bibnamefont {Tufarelli}},
  \bibinfo {author} {\bibfnamefont {M.}~\bibnamefont {Piano}}, \ and\ \bibinfo
  {author} {\bibfnamefont {G.}~\bibnamefont {Adesso}},\ }\href@noop {}
  {\bibfield  {journal} {\bibinfo  {journal} {Phys. Rev. Lett.}\ }\textbf
  {\bibinfo {volume} {121}},\ \bibinfo {pages} {160401} (\bibinfo {year}
  {2018})}\BibitemShut {NoStop}%
\bibitem [{\citenamefont {Brandes}(2010)}]{Brandes2010}%
  \BibitemOpen
  \bibfield  {author} {\bibinfo {author} {\bibfnamefont {T.}~\bibnamefont
  {Brandes}},\ }\href@noop {} {\bibfield  {journal} {\bibinfo  {journal} {Phys.
  Rev. Lett.}\ }\textbf {\bibinfo {volume} {105}},\ \bibinfo {pages} {060602}
  (\bibinfo {year} {2010})}\BibitemShut {NoStop}%
\bibitem [{\citenamefont {Kiesslich}\ \emph {et~al.}(2012)\citenamefont
  {Kiesslich}, \citenamefont {Emary}, \citenamefont {Schaller},\ and\
  \citenamefont {Brandes}}]{Kiesslich2012}%
  \BibitemOpen
  \bibfield  {author} {\bibinfo {author} {\bibfnamefont {G.}~\bibnamefont
  {Kiesslich}}, \bibinfo {author} {\bibfnamefont {C.}~\bibnamefont {Emary}},
  \bibinfo {author} {\bibfnamefont {G.}~\bibnamefont {Schaller}}, \ and\
  \bibinfo {author} {\bibfnamefont {T.}~\bibnamefont {Brandes}},\ }\href@noop
  {} {\bibfield  {journal} {\bibinfo  {journal} {New J. Phys.}\ }\textbf
  {\bibinfo {volume} {14}},\ \bibinfo {pages} {123036} (\bibinfo {year}
  {2012})}\BibitemShut {NoStop}%
\bibitem [{\citenamefont {Gough}(2014)}]{Gough2014}%
  \BibitemOpen
  \bibfield  {author} {\bibinfo {author} {\bibfnamefont {J.}~\bibnamefont
  {Gough}},\ }\href@noop {} {\bibfield  {journal} {\bibinfo  {journal} {Phys.
  Rev. E}\ }\textbf {\bibinfo {volume} {90}},\ \bibinfo {pages} {062109}
  (\bibinfo {year} {2014})}\BibitemShut {NoStop}%
\bibitem [{\citenamefont {Brandes}\ and\ \citenamefont
  {Emary}(2016)}]{Brandes2016}%
  \BibitemOpen
  \bibfield  {author} {\bibinfo {author} {\bibfnamefont {T.}~\bibnamefont
  {Brandes}}\ and\ \bibinfo {author} {\bibfnamefont {C.}~\bibnamefont
  {Emary}},\ }\href@noop {} {\bibfield  {journal} {\bibinfo  {journal} {Phys.
  Rev. E}\ }\textbf {\bibinfo {volume} {93}},\ \bibinfo {pages} {042103}
  (\bibinfo {year} {2016})}\BibitemShut {NoStop}%
\bibitem [{\citenamefont {Luo}\ \emph {et~al.}(2016)\citenamefont {Luo},
  \citenamefont {Jin}, \citenamefont {Wang}, \citenamefont {Hu}, \citenamefont
  {Huang},\ and\ \citenamefont {He}}]{Luo2016}%
  \BibitemOpen
  \bibfield  {author} {\bibinfo {author} {\bibfnamefont {J.~Y.}\ \bibnamefont
  {Luo}}, \bibinfo {author} {\bibfnamefont {J.}~\bibnamefont {Jin}}, \bibinfo
  {author} {\bibfnamefont {S.-K.}\ \bibnamefont {Wang}}, \bibinfo {author}
  {\bibfnamefont {J.}~\bibnamefont {Hu}}, \bibinfo {author} {\bibfnamefont
  {Y.}~\bibnamefont {Huang}}, \ and\ \bibinfo {author} {\bibfnamefont {X.-L.}\
  \bibnamefont {He}},\ }\href@noop {} {\bibfield  {journal} {\bibinfo
  {journal} {Phys. Rev. B}\ }\textbf {\bibinfo {volume} {93}},\ \bibinfo
  {pages} {125122} (\bibinfo {year} {2016})}\BibitemShut {NoStop}%
\bibitem [{\citenamefont {Wagner}\ \emph {et~al.}(2017)\citenamefont {Wagner},
  \citenamefont {Strasberg}, \citenamefont {Bayer}, \citenamefont
  {Rugeramigabo}, \citenamefont {Brandes},\ and\ \citenamefont
  {Haug}}]{Wagner2016}%
  \BibitemOpen
  \bibfield  {author} {\bibinfo {author} {\bibfnamefont {T.}~\bibnamefont
  {Wagner}}, \bibinfo {author} {\bibfnamefont {P.}~\bibnamefont {Strasberg}},
  \bibinfo {author} {\bibfnamefont {J.~C.}\ \bibnamefont {Bayer}}, \bibinfo
  {author} {\bibfnamefont {E.~P.}\ \bibnamefont {Rugeramigabo}}, \bibinfo
  {author} {\bibfnamefont {T.}~\bibnamefont {Brandes}}, \ and\ \bibinfo
  {author} {\bibfnamefont {R.~J.}\ \bibnamefont {Haug}},\ }\href@noop {}
  {\bibfield  {journal} {\bibinfo  {journal} {Nat. Nanotechnol.}\ }\textbf
  {\bibinfo {volume} {12}},\ \bibinfo {pages} {218} (\bibinfo {year}
  {2017})}\BibitemShut {NoStop}%
\bibitem [{\citenamefont {Beige}\ \emph {et~al.}(2000)\citenamefont {Beige},
  \citenamefont {Braun}, \citenamefont {Tregenna},\ and\ \citenamefont
  {Knight}}]{Beige2000}%
  \BibitemOpen
  \bibfield  {author} {\bibinfo {author} {\bibfnamefont {A.}~\bibnamefont
  {Beige}}, \bibinfo {author} {\bibfnamefont {D.}~\bibnamefont {Braun}},
  \bibinfo {author} {\bibfnamefont {B.}~\bibnamefont {Tregenna}}, \ and\
  \bibinfo {author} {\bibfnamefont {P.~L.}\ \bibnamefont {Knight}},\
  }\href@noop {} {\bibfield  {journal} {\bibinfo  {journal} {Phys. Rev. Lett.}\
  }\textbf {\bibinfo {volume} {85}},\ \bibinfo {pages} {1762} (\bibinfo {year}
  {2000})}\BibitemShut {NoStop}%
\bibitem [{\citenamefont {Verstraete}\ \emph {et~al.}(2009)\citenamefont
  {Verstraete}, \citenamefont {Wolf},\ and\ \citenamefont
  {Cirac}}]{Verstraete2009}%
  \BibitemOpen
  \bibfield  {author} {\bibinfo {author} {\bibfnamefont {F.}~\bibnamefont
  {Verstraete}}, \bibinfo {author} {\bibfnamefont {M.~M.}\ \bibnamefont
  {Wolf}}, \ and\ \bibinfo {author} {\bibfnamefont {J.~I.}\ \bibnamefont
  {Cirac}},\ }\href@noop {} {\bibfield  {journal} {\bibinfo  {journal} {Nat.
  Phys}\ }\textbf {\bibinfo {volume} {5}},\ \bibinfo {pages} {633} (\bibinfo
  {year} {2009})}\BibitemShut {NoStop}%
\bibitem [{\citenamefont {Zanardi}\ \emph {et~al.}(2016)\citenamefont
  {Zanardi}, \citenamefont {Marshall},\ and\ \citenamefont
  {Venuti}}]{Zanardi2016}%
  \BibitemOpen
  \bibfield  {author} {\bibinfo {author} {\bibfnamefont {P.}~\bibnamefont
  {Zanardi}}, \bibinfo {author} {\bibfnamefont {J.}~\bibnamefont {Marshall}}, \
  and\ \bibinfo {author} {\bibfnamefont {L.~C.}\ \bibnamefont {Venuti}},\
  }\href@noop {} {\bibfield  {journal} {\bibinfo  {journal} {Phys. Rev. A}\
  }\textbf {\bibinfo {volume} {93}},\ \bibinfo {pages} {022312} (\bibinfo
  {year} {2016})}\BibitemShut {NoStop}%
\bibitem [{\citenamefont {Kapit}(2018)}]{Kapit2018}%
  \BibitemOpen
  \bibfield  {author} {\bibinfo {author} {\bibfnamefont {E.}~\bibnamefont
  {Kapit}},\ }\href@noop {} {\bibfield  {journal} {\bibinfo  {journal} {Phys.
  Rev. Lett.}\ }\textbf {\bibinfo {volume} {120}},\ \bibinfo {pages} {050503}
  (\bibinfo {year} {2018})}\BibitemShut {NoStop}%
\bibitem [{\citenamefont {de~Vega}\ and\ \citenamefont
  {Alonso}(2017)}]{deVega2017}%
  \BibitemOpen
  \bibfield  {author} {\bibinfo {author} {\bibfnamefont {I.}~\bibnamefont
  {de~Vega}}\ and\ \bibinfo {author} {\bibfnamefont {D.}~\bibnamefont
  {Alonso}},\ }\href@noop {} {\bibfield  {journal} {\bibinfo  {journal} {Rev.
  Mod. Phys.}\ }\textbf {\bibinfo {volume} {89}},\ \bibinfo {pages} {015001}
  (\bibinfo {year} {2017})}\BibitemShut {NoStop}%
\bibitem [{\citenamefont {Pruschke}\ \emph {et~al.}(1995)\citenamefont
  {Pruschke}, \citenamefont {Jarrell},\ and\ \citenamefont
  {Freericks}}]{Pruschke1995}%
  \BibitemOpen
  \bibfield  {author} {\bibinfo {author} {\bibfnamefont {T.}~\bibnamefont
  {Pruschke}}, \bibinfo {author} {\bibfnamefont {M.}~\bibnamefont {Jarrell}}, \
  and\ \bibinfo {author} {\bibfnamefont {J.}~\bibnamefont {Freericks}},\ }\href
  {\doibase 10.1080/00018739500101526} {\bibfield  {journal} {\bibinfo
  {journal} {Advances in Physics}\ }\textbf {\bibinfo {volume} {44}},\ \bibinfo
  {pages} {187} (\bibinfo {year} {1995})},\ \Eprint
  {http://arxiv.org/abs/https://doi.org/10.1080/00018739500101526}
  {https://doi.org/10.1080/00018739500101526} \BibitemShut {NoStop}%
\bibitem [{\citenamefont {Georges}\ \emph {et~al.}(1996)\citenamefont
  {Georges}, \citenamefont {Kotliar}, \citenamefont {Krauth},\ and\
  \citenamefont {Rozenberg}}]{Georges1996}%
  \BibitemOpen
  \bibfield  {author} {\bibinfo {author} {\bibfnamefont {A.}~\bibnamefont
  {Georges}}, \bibinfo {author} {\bibfnamefont {G.}~\bibnamefont {Kotliar}},
  \bibinfo {author} {\bibfnamefont {W.}~\bibnamefont {Krauth}}, \ and\ \bibinfo
  {author} {\bibfnamefont {M.~J.}\ \bibnamefont {Rozenberg}},\ }\href {\doibase
  10.1103/RevModPhys.68.13} {\bibfield  {journal} {\bibinfo  {journal} {Rev.
  Mod. Phys.}\ }\textbf {\bibinfo {volume} {68}},\ \bibinfo {pages} {13}
  (\bibinfo {year} {1996})}\BibitemShut {NoStop}%
\bibitem [{\citenamefont {Freericks}\ \emph {et~al.}(2006)\citenamefont
  {Freericks}, \citenamefont {Turkowski},\ and\ \citenamefont
  {Zlatic}}]{Freericks2006}%
  \BibitemOpen
  \bibfield  {author} {\bibinfo {author} {\bibfnamefont {J.~K.}\ \bibnamefont
  {Freericks}}, \bibinfo {author} {\bibfnamefont {V.~M.}\ \bibnamefont
  {Turkowski}}, \ and\ \bibinfo {author} {\bibfnamefont {V.}~\bibnamefont
  {Zlatic}},\ }\href@noop {} {\bibfield  {journal} {\bibinfo  {journal} {Phys.
  Rev. Lett.}\ }\textbf {\bibinfo {volume} {97}},\ \bibinfo {pages} {266408}
  (\bibinfo {year} {2006})}\BibitemShut {NoStop}%
\bibitem [{\citenamefont {Aoki}\ \emph {et~al.}(2014)\citenamefont {Aoki},
  \citenamefont {Tsuji}, \citenamefont {Eckstein}, \citenamefont {Kollar},
  \citenamefont {Oka},\ and\ \citenamefont {Werner}}]{Aoki2014}%
  \BibitemOpen
  \bibfield  {author} {\bibinfo {author} {\bibfnamefont {H.}~\bibnamefont
  {Aoki}}, \bibinfo {author} {\bibfnamefont {N.}~\bibnamefont {Tsuji}},
  \bibinfo {author} {\bibfnamefont {M.}~\bibnamefont {Eckstein}}, \bibinfo
  {author} {\bibfnamefont {M.}~\bibnamefont {Kollar}}, \bibinfo {author}
  {\bibfnamefont {T.}~\bibnamefont {Oka}}, \ and\ \bibinfo {author}
  {\bibfnamefont {P.}~\bibnamefont {Werner}},\ }\href@noop {} {\bibfield
  {journal} {\bibinfo  {journal} {Rev. Mod. Phys.}\ }\textbf {\bibinfo {volume}
  {86}},\ \bibinfo {pages} {779} (\bibinfo {year} {2014})}\BibitemShut
  {NoStop}%
\bibitem [{\citenamefont {Breuer}\ \emph {et~al.}(2016)\citenamefont {Breuer},
  \citenamefont {Laine}, \citenamefont {Piilo},\ and\ \citenamefont
  {Vacchini}}]{Breuer2016}%
  \BibitemOpen
  \bibfield  {author} {\bibinfo {author} {\bibfnamefont {H.-P.}\ \bibnamefont
  {Breuer}}, \bibinfo {author} {\bibfnamefont {E.-M.}\ \bibnamefont {Laine}},
  \bibinfo {author} {\bibfnamefont {J.}~\bibnamefont {Piilo}}, \ and\ \bibinfo
  {author} {\bibfnamefont {B.}~\bibnamefont {Vacchini}},\ }\href@noop {}
  {\bibfield  {journal} {\bibinfo  {journal} {Rev. Mod. Phys.}\ }\textbf
  {\bibinfo {volume} {88}},\ \bibinfo {pages} {021002} (\bibinfo {year}
  {2016})}\BibitemShut {NoStop}%
\bibitem [{\citenamefont {Wiseman}\ and\ \citenamefont
  {Milburn}(1993)}]{Wiseman1993}%
  \BibitemOpen
  \bibfield  {author} {\bibinfo {author} {\bibfnamefont {H.~M.}\ \bibnamefont
  {Wiseman}}\ and\ \bibinfo {author} {\bibfnamefont {G.~J.}\ \bibnamefont
  {Milburn}},\ }\href@noop {} {\bibfield  {journal} {\bibinfo  {journal} {Phys.
  Rev. A}\ }\textbf {\bibinfo {volume} {47}},\ \bibinfo {pages} {642} (\bibinfo
  {year} {1993})}\BibitemShut {NoStop}%
\bibitem [{\citenamefont {Wiseman}(1996)}]{Wiseman1996}%
  \BibitemOpen
  \bibfield  {author} {\bibinfo {author} {\bibfnamefont {H.~M.}\ \bibnamefont
  {Wiseman}},\ }\href@noop {} {\bibfield  {journal} {\bibinfo  {journal}
  {Quantum Semiclass. Opt.}\ }\textbf {\bibinfo {volume} {8}},\ \bibinfo
  {pages} {205} (\bibinfo {year} {1996})}\BibitemShut {NoStop}%
\bibitem [{\citenamefont {Percival}(1999)}]{Percival1999}%
  \BibitemOpen
  \bibfield  {author} {\bibinfo {author} {\bibfnamefont {I.}~\bibnamefont
  {Percival}},\ }\href@noop {} {\emph {\bibinfo {title} {Quantum State
  Diffusion}}}\ (\bibinfo  {publisher} {Cambridge University Press},\ \bibinfo
  {address} {Cambridge},\ \bibinfo {year} {1999})\BibitemShut {NoStop}%
\bibitem [{\citenamefont {Warszawski}\ and\ \citenamefont
  {Wiseman}(2003{\natexlab{a}})}]{Warszawski2003}%
  \BibitemOpen
  \bibfield  {author} {\bibinfo {author} {\bibfnamefont {P.}~\bibnamefont
  {Warszawski}}\ and\ \bibinfo {author} {\bibfnamefont {H.~M.}\ \bibnamefont
  {Wiseman}},\ }\href@noop {} {\bibfield  {journal} {\bibinfo  {journal} {J.
  Opt. B: Quantum Semiclass. Opt.}\ }\textbf {\bibinfo {volume} {5}},\ \bibinfo
  {pages} {1} (\bibinfo {year} {2003}{\natexlab{a}})}\BibitemShut {NoStop}%
\bibitem [{\citenamefont {Warszawski}\ and\ \citenamefont
  {Wiseman}(2003{\natexlab{b}})}]{Warszawski2003a}%
  \BibitemOpen
  \bibfield  {author} {\bibinfo {author} {\bibfnamefont {P.}~\bibnamefont
  {Warszawski}}\ and\ \bibinfo {author} {\bibfnamefont {H.~M.}\ \bibnamefont
  {Wiseman}},\ }\href@noop {} {\bibfield  {journal} {\bibinfo  {journal} {J.
  Opt. B: Quantum Semiclass. Opt.}\ }\textbf {\bibinfo {volume} {5}},\ \bibinfo
  {pages} {15} (\bibinfo {year} {2003}{\natexlab{b}})}\BibitemShut {NoStop}%
\bibitem [{\citenamefont {Oxtoby}\ \emph {et~al.}(2005)\citenamefont {Oxtoby},
  \citenamefont {Warszawski}, \citenamefont {Wiseman}, \citenamefont {Sun},\
  and\ \citenamefont {Polkinghorne}}]{Oxtoby2005}%
  \BibitemOpen
  \bibfield  {author} {\bibinfo {author} {\bibfnamefont {N.~P.}\ \bibnamefont
  {Oxtoby}}, \bibinfo {author} {\bibfnamefont {P.}~\bibnamefont {Warszawski}},
  \bibinfo {author} {\bibfnamefont {H.~M.}\ \bibnamefont {Wiseman}}, \bibinfo
  {author} {\bibfnamefont {H.-B.}\ \bibnamefont {Sun}}, \ and\ \bibinfo
  {author} {\bibfnamefont {R.~E.~S.}\ \bibnamefont {Polkinghorne}},\
  }\href@noop {} {\bibfield  {journal} {\bibinfo  {journal} {Phys. Rev. B}\
  }\textbf {\bibinfo {volume} {71}},\ \bibinfo {pages} {165317} (\bibinfo
  {year} {2005})}\BibitemShut {NoStop}%
\bibitem [{\citenamefont {Tilloy}\ \emph {et~al.}(2015)\citenamefont {Tilloy},
  \citenamefont {Bauer},\ and\ \citenamefont {Bernard}}]{Tilloy2015}%
  \BibitemOpen
  \bibfield  {author} {\bibinfo {author} {\bibfnamefont {A.}~\bibnamefont
  {Tilloy}}, \bibinfo {author} {\bibfnamefont {M.}~\bibnamefont {Bauer}}, \
  and\ \bibinfo {author} {\bibfnamefont {D.}~\bibnamefont {Bernard}},\
  }\href@noop {} {\bibfield  {journal} {\bibinfo  {journal} {Phys. Rev. A}\
  }\textbf {\bibinfo {volume} {92}},\ \bibinfo {pages} {052111} (\bibinfo
  {year} {2015})}\BibitemShut {NoStop}%
\bibitem [{\citenamefont {Bauer}\ \emph {et~al.}(2015)\citenamefont {Bauer},
  \citenamefont {Bernard},\ and\ \citenamefont {Tilloy}}]{Bauer2015}%
  \BibitemOpen
  \bibfield  {author} {\bibinfo {author} {\bibfnamefont {M.}~\bibnamefont
  {Bauer}}, \bibinfo {author} {\bibfnamefont {D.}~\bibnamefont {Bernard}}, \
  and\ \bibinfo {author} {\bibfnamefont {A.}~\bibnamefont {Tilloy}},\
  }\href@noop {} {\bibfield  {journal} {\bibinfo  {journal} {J. Phys. A: Math.
  Theor.}\ }\textbf {\bibinfo {volume} {48}},\ \bibinfo {pages} {25FT02}
  (\bibinfo {year} {2015})}\BibitemShut {NoStop}%
\bibitem [{\citenamefont {Daley}(2014)}]{Daley2014}%
  \BibitemOpen
  \bibfield  {author} {\bibinfo {author} {\bibfnamefont {A.~J.}\ \bibnamefont
  {Daley}},\ }\href@noop {} {\bibfield  {journal} {\bibinfo  {journal} {Adv. in
  Phys.}\ }\textbf {\bibinfo {volume} {63}},\ \bibinfo {pages} {77} (\bibinfo
  {year} {2014})}\BibitemShut {NoStop}%
\bibitem [{\citenamefont {Zhang}\ \emph {et~al.}(2017)\citenamefont {Zhang},
  \citenamefont {Liu}, \citenamefont {Wu}, \citenamefont {Jacobs},\ and\
  \citenamefont {Nori}}]{Zhang2017}%
  \BibitemOpen
  \bibfield  {author} {\bibinfo {author} {\bibfnamefont {J.}~\bibnamefont
  {Zhang}}, \bibinfo {author} {\bibfnamefont {Y.-x.}\ \bibnamefont {Liu}},
  \bibinfo {author} {\bibfnamefont {R.-B.}\ \bibnamefont {Wu}}, \bibinfo
  {author} {\bibfnamefont {K.}~\bibnamefont {Jacobs}}, \ and\ \bibinfo {author}
  {\bibfnamefont {F.}~\bibnamefont {Nori}},\ }\href@noop {} {\bibfield
  {journal} {\bibinfo  {journal} {Physics Reports}\ }\textbf {\bibinfo {volume}
  {679}},\ \bibinfo {pages} {1} (\bibinfo {year} {2017})}\BibitemShut {NoStop}%
\bibitem [{\citenamefont {Gardiner}\ and\ \citenamefont
  {Collett}(1985)}]{Gardiner1985}%
  \BibitemOpen
  \bibfield  {author} {\bibinfo {author} {\bibfnamefont {C.~W.}\ \bibnamefont
  {Gardiner}}\ and\ \bibinfo {author} {\bibfnamefont {M.~J.}\ \bibnamefont
  {Collett}},\ }\href@noop {} {\bibfield  {journal} {\bibinfo  {journal} {Phys.
  Rev. A}\ }\textbf {\bibinfo {volume} {31}},\ \bibinfo {pages} {3761}
  (\bibinfo {year} {1985})}\BibitemShut {NoStop}%
\bibitem [{\citenamefont {Gardiner}\ and\ \citenamefont
  {Parkins}(1987)}]{Gardiner1987}%
  \BibitemOpen
  \bibfield  {author} {\bibinfo {author} {\bibfnamefont {C.~W.}\ \bibnamefont
  {Gardiner}}\ and\ \bibinfo {author} {\bibfnamefont {A.~S.}\ \bibnamefont
  {Parkins}},\ }\href@noop {} {\bibfield  {journal} {\bibinfo  {journal} {J.
  Opt. Soc. Am. B}\ }\textbf {\bibinfo {volume} {4}},\ \bibinfo {pages} {1683}
  (\bibinfo {year} {1987})}\BibitemShut {NoStop}%
\bibitem [{\citenamefont {Gardiner}(2004)}]{Gardiner2004}%
  \BibitemOpen
  \bibfield  {author} {\bibinfo {author} {\bibfnamefont {C.~W.}\ \bibnamefont
  {Gardiner}},\ }\href@noop {} {\bibfield  {journal} {\bibinfo  {journal} {Opt.
  Commun}\ }\textbf {\bibinfo {volume} {243}},\ \bibinfo {pages} {57} (\bibinfo
  {year} {2004})}\BibitemShut {NoStop}%
\bibitem [{\citenamefont {Li}\ \emph {et~al.}(2018)\citenamefont {Li},
  \citenamefont {Michael},\ and\ \citenamefont {Wiseman}}]{Li2018}%
  \BibitemOpen
  \bibfield  {author} {\bibinfo {author} {\bibfnamefont {L.}~\bibnamefont
  {Li}}, \bibinfo {author} {\bibfnamefont {J.~W.~H.}\ \bibnamefont {Michael}},
  \ and\ \bibinfo {author} {\bibfnamefont {H.~M.}\ \bibnamefont {Wiseman}},\
  }\href@noop {} {\bibfield  {journal} {\bibinfo  {journal} {Phys. Rep.}\
  }\textbf {\bibinfo {volume} {759}},\ \bibinfo {pages} {1} (\bibinfo {year}
  {2018})}\BibitemShut {NoStop}%
\bibitem [{\citenamefont {Jack}\ \emph {et~al.}(1999)\citenamefont {Jack},
  \citenamefont {Collett},\ and\ \citenamefont {Walls}}]{Jack1999}%
  \BibitemOpen
  \bibfield  {author} {\bibinfo {author} {\bibfnamefont {M.~W.}\ \bibnamefont
  {Jack}}, \bibinfo {author} {\bibfnamefont {M.~J.}\ \bibnamefont {Collett}}, \
  and\ \bibinfo {author} {\bibfnamefont {D.~F.}\ \bibnamefont {Walls}},\
  }\href@noop {} {\bibfield  {journal} {\bibinfo  {journal} {J. Opt. B: Quantum
  Semiclass. Opt.}\ }\textbf {\bibinfo {volume} {1}},\ \bibinfo {pages} {452}
  (\bibinfo {year} {1999})}\BibitemShut {NoStop}%
\bibitem [{\citenamefont {Xu}\ and\ \citenamefont {Li}(2018)}]{Xu2018}%
  \BibitemOpen
  \bibfield  {author} {\bibinfo {author} {\bibfnamefont {L.}~\bibnamefont
  {Xu}}\ and\ \bibinfo {author} {\bibfnamefont {X.-Q.}\ \bibnamefont {Li}},\
  }\href@noop {} {\bibfield  {journal} {\bibinfo  {journal} {Sci. Rep.}\
  }\textbf {\bibinfo {volume} {8}},\ \bibinfo {pages} {452} (\bibinfo {year}
  {2018})}\BibitemShut {NoStop}%
\bibitem [{\citenamefont {Breuer}\ and\ \citenamefont
  {Petruccione}(2007)}]{Breuer2011}%
  \BibitemOpen
  \bibfield  {author} {\bibinfo {author} {\bibfnamefont {H.-P.}\ \bibnamefont
  {Breuer}}\ and\ \bibinfo {author} {\bibfnamefont {F.}~\bibnamefont
  {Petruccione}},\ }\href@noop {} {\emph {\bibinfo {title} {The Theory of Open
  Quantum Systems}}}\ (\bibinfo  {publisher} {Oxford University Press},\
  \bibinfo {address} {Oxford},\ \bibinfo {year} {2007})\BibitemShut {NoStop}%
\bibitem [{\citenamefont {Diosi}(2012)}]{Diosi2012}%
  \BibitemOpen
  \bibfield  {author} {\bibinfo {author} {\bibfnamefont {L.}~\bibnamefont
  {Diosi}},\ }\href@noop {} {\bibfield  {journal} {\bibinfo  {journal} {Phys.
  Rev. A}\ }\textbf {\bibinfo {volume} {85}},\ \bibinfo {pages} {034101}
  (\bibinfo {year} {2012})}\BibitemShut {NoStop}%
\bibitem [{\citenamefont {Garraway}(1997)}]{Garraway1996}%
  \BibitemOpen
  \bibfield  {author} {\bibinfo {author} {\bibfnamefont {B.~M.}\ \bibnamefont
  {Garraway}},\ }\href@noop {} {\bibfield  {journal} {\bibinfo  {journal}
  {Phys. Rev. A}\ }\textbf {\bibinfo {volume} {55}},\ \bibinfo {pages} {2290}
  (\bibinfo {year} {1997})}\BibitemShut {NoStop}%
\bibitem [{\citenamefont {Mazzola}\ \emph {et~al.}(2009)\citenamefont
  {Mazzola}, \citenamefont {Maniscalco}, \citenamefont {Piilo}, \citenamefont
  {Suominen},\ and\ \citenamefont {Garraway}}]{Mazzola2009}%
  \BibitemOpen
  \bibfield  {author} {\bibinfo {author} {\bibfnamefont {L.}~\bibnamefont
  {Mazzola}}, \bibinfo {author} {\bibfnamefont {S.}~\bibnamefont {Maniscalco}},
  \bibinfo {author} {\bibfnamefont {J.}~\bibnamefont {Piilo}}, \bibinfo
  {author} {\bibfnamefont {K.-A.}\ \bibnamefont {Suominen}}, \ and\ \bibinfo
  {author} {\bibfnamefont {B.~M.}\ \bibnamefont {Garraway}},\ }\href@noop {}
  {\bibfield  {journal} {\bibinfo  {journal} {Phys. Rev. A}\ }\textbf {\bibinfo
  {volume} {80}},\ \bibinfo {pages} {012104} (\bibinfo {year}
  {2009})}\BibitemShut {NoStop}%
\bibitem [{\citenamefont {Arrigoni}\ \emph {et~al.}(2013)\citenamefont
  {Arrigoni}, \citenamefont {Knap},\ and\ \citenamefont
  {Linden}}]{Arrigoni2013}%
  \BibitemOpen
  \bibfield  {author} {\bibinfo {author} {\bibfnamefont {E.}~\bibnamefont
  {Arrigoni}}, \bibinfo {author} {\bibfnamefont {M.}~\bibnamefont {Knap}}, \
  and\ \bibinfo {author} {\bibfnamefont {W.~v.~d.}\ \bibnamefont {Linden}},\
  }\href@noop {} {\bibfield  {journal} {\bibinfo  {journal} {Phys. Rev. Lett.}\
  }\textbf {\bibinfo {volume} {110}},\ \bibinfo {pages} {086403} (\bibinfo
  {year} {2013})}\BibitemShut {NoStop}%
\bibitem [{\citenamefont {Budini}(2013)}]{Budini2013}%
  \BibitemOpen
  \bibfield  {author} {\bibinfo {author} {\bibfnamefont {A.~A.}\ \bibnamefont
  {Budini}},\ }\href@noop {} {\bibfield  {journal} {\bibinfo  {journal} {Phys.
  Rev. A}\ }\textbf {\bibinfo {volume} {88}},\ \bibinfo {pages} {012124}
  (\bibinfo {year} {2013})}\BibitemShut {NoStop}%
\bibitem [{\citenamefont {Dorda}\ \emph {et~al.}(2014)\citenamefont {Dorda},
  \citenamefont {Linden},\ and\ \citenamefont {Arrigoni}}]{Dorda2014}%
  \BibitemOpen
  \bibfield  {author} {\bibinfo {author} {\bibfnamefont {M.}~\bibnamefont
  {Dorda}, \bibfnamefont {Antonius amd~Nuss}}, \bibinfo {author} {\bibfnamefont
  {W.~v.~d.}\ \bibnamefont {Linden}}, \ and\ \bibinfo {author} {\bibfnamefont
  {E.}~\bibnamefont {Arrigoni}},\ }\href@noop {} {\bibfield  {journal}
  {\bibinfo  {journal} {Phys. Rev. B}\ }\textbf {\bibinfo {volume} {89}},\
  \bibinfo {pages} {165105} (\bibinfo {year} {2014})}\BibitemShut {NoStop}%
\bibitem [{\citenamefont {Strathearn}\ \emph {et~al.}(2018)\citenamefont
  {Strathearn}, \citenamefont {Kirton}, \citenamefont {Kilda}, \citenamefont
  {Keeling},\ and\ \citenamefont {Lovett}}]{Strathearn2018}%
  \BibitemOpen
  \bibfield  {author} {\bibinfo {author} {\bibfnamefont {A.}~\bibnamefont
  {Strathearn}}, \bibinfo {author} {\bibfnamefont {P.}~\bibnamefont {Kirton}},
  \bibinfo {author} {\bibfnamefont {D.}~\bibnamefont {Kilda}}, \bibinfo
  {author} {\bibfnamefont {J.}~\bibnamefont {Keeling}}, \ and\ \bibinfo
  {author} {\bibfnamefont {B.~W.}\ \bibnamefont {Lovett}},\ }\href@noop {}
  {\bibfield  {journal} {\bibinfo  {journal} {Nat. Commun.}\ }\textbf {\bibinfo
  {volume} {9}},\ \bibinfo {pages} {3322} (\bibinfo {year} {2018})}\BibitemShut
  {NoStop}%
\bibitem [{\citenamefont {Makarov}\ and\ \citenamefont
  {Makri}(1994)}]{Makarov1994}%
  \BibitemOpen
  \bibfield  {author} {\bibinfo {author} {\bibfnamefont {D.~E.}\ \bibnamefont
  {Makarov}}\ and\ \bibinfo {author} {\bibfnamefont {N.}~\bibnamefont
  {Makri}},\ }\href {\doibase https://doi.org/10.1016/0009-2614(94)00275-4}
  {\bibfield  {journal} {\bibinfo  {journal} {Chemical Physics Letters}\
  }\textbf {\bibinfo {volume} {221}},\ \bibinfo {pages} {482 } (\bibinfo {year}
  {1994})}\BibitemShut {NoStop}%
\bibitem [{\citenamefont {Makri}(1995)}]{Makri1995}%
  \BibitemOpen
  \bibfield  {author} {\bibinfo {author} {\bibfnamefont {N.}~\bibnamefont
  {Makri}},\ }\href {\doibase 10.1063/1.531046} {\bibfield  {journal} {\bibinfo
   {journal} {Journal of Mathematical Physics}\ }\textbf {\bibinfo {volume}
  {36}},\ \bibinfo {pages} {2430} (\bibinfo {year} {1995})},\ \Eprint
  {http://arxiv.org/abs/https://doi.org/10.1063/1.531046}
  {https://doi.org/10.1063/1.531046} \BibitemShut {NoStop}%
\bibitem [{\citenamefont {Makri}\ and\ \citenamefont
  {Makarov}(1995)}]{Makri1995a}%
  \BibitemOpen
  \bibfield  {author} {\bibinfo {author} {\bibfnamefont {N.}~\bibnamefont
  {Makri}}\ and\ \bibinfo {author} {\bibfnamefont {D.~E.}\ \bibnamefont
  {Makarov}},\ }\href {\doibase 10.1063/1.469508} {\bibfield  {journal}
  {\bibinfo  {journal} {The Journal of Chemical Physics}\ }\textbf {\bibinfo
  {volume} {102}},\ \bibinfo {pages} {4600} (\bibinfo {year} {1995})},\ \Eprint
  {http://arxiv.org/abs/https://doi.org/10.1063/1.469508}
  {https://doi.org/10.1063/1.469508} \BibitemShut {NoStop}%
\bibitem [{\citenamefont {Makri}\ \emph {et~al.}(1996)\citenamefont {Makri},
  \citenamefont {Sim}, \citenamefont {Makarov},\ and\ \citenamefont
  {Topaler}}]{Makri1996}%
  \BibitemOpen
  \bibfield  {author} {\bibinfo {author} {\bibfnamefont {N.}~\bibnamefont
  {Makri}}, \bibinfo {author} {\bibfnamefont {E.}~\bibnamefont {Sim}}, \bibinfo
  {author} {\bibfnamefont {D.~E.}\ \bibnamefont {Makarov}}, \ and\ \bibinfo
  {author} {\bibfnamefont {M.}~\bibnamefont {Topaler}},\ }\href
  {http://www.pnas.org/content/93/9/3926.abstract} {\bibfield  {journal}
  {\bibinfo  {journal} {Proceedings of the National Academy of Sciences}\
  }\textbf {\bibinfo {volume} {93}},\ \bibinfo {pages} {3926} (\bibinfo {year}
  {1996})},\ \Eprint
  {http://arxiv.org/abs/http://www.pnas.org/content/93/9/3926.full.pdf}
  {http://www.pnas.org/content/93/9/3926.full.pdf} \BibitemShut {NoStop}%
\bibitem [{\citenamefont {Diosi}\ \emph {et~al.}(1998)\citenamefont {Diosi},
  \citenamefont {Gisin},\ and\ \citenamefont {Strunz}}]{Diosi1998}%
  \BibitemOpen
  \bibfield  {author} {\bibinfo {author} {\bibfnamefont {L.}~\bibnamefont
  {Diosi}}, \bibinfo {author} {\bibfnamefont {N.}~\bibnamefont {Gisin}}, \ and\
  \bibinfo {author} {\bibfnamefont {W.~T.}\ \bibnamefont {Strunz}},\
  }\href@noop {} {\bibfield  {journal} {\bibinfo  {journal} {Phys.l Rev. A}\
  }\textbf {\bibinfo {volume} {58}},\ \bibinfo {pages} {1699} (\bibinfo {year}
  {1998})}\BibitemShut {NoStop}%
\bibitem [{\citenamefont {Suess}\ \emph {et~al.}(2014)\citenamefont {Suess},
  \citenamefont {Eisfeld},\ and\ \citenamefont {Strunz}}]{Suess2014}%
  \BibitemOpen
  \bibfield  {author} {\bibinfo {author} {\bibfnamefont {D.}~\bibnamefont
  {Suess}}, \bibinfo {author} {\bibfnamefont {A.}~\bibnamefont {Eisfeld}}, \
  and\ \bibinfo {author} {\bibfnamefont {W.~T.}\ \bibnamefont {Strunz}},\
  }\href@noop {} {\bibfield  {journal} {\bibinfo  {journal} {Phys. Rev. Lett.}\
  }\textbf {\bibinfo {volume} {113}},\ \bibinfo {pages} {150403} (\bibinfo
  {year} {2014})}\BibitemShut {NoStop}%
\bibitem [{\citenamefont {Hartmann}\ and\ \citenamefont
  {Strunz}(2017)}]{Hartmann2017}%
  \BibitemOpen
  \bibfield  {author} {\bibinfo {author} {\bibfnamefont {R.}~\bibnamefont
  {Hartmann}}\ and\ \bibinfo {author} {\bibfnamefont {W.~T.}\ \bibnamefont
  {Strunz}},\ }\href@noop {} {\bibfield  {journal} {\bibinfo  {journal} {J.
  Chem. Theor. Comput.}\ }\textbf {\bibinfo {volume} {13}},\ \bibinfo {pages}
  {5834} (\bibinfo {year} {2017})}\BibitemShut {NoStop}%
\bibitem [{\citenamefont {Shao}(2004)}]{Shao2004}%
  \BibitemOpen
  \bibfield  {author} {\bibinfo {author} {\bibfnamefont {J.}~\bibnamefont
  {Shao}},\ }\href@noop {} {\bibfield  {journal} {\bibinfo  {journal} {J. Chem.
  Phys.}\ }\textbf {\bibinfo {volume} {120}},\ \bibinfo {pages} {5053}
  (\bibinfo {year} {2004})}\BibitemShut {NoStop}%
\bibitem [{\citenamefont {Yan}\ \emph {et~al.}(2004)\citenamefont {Yan},
  \citenamefont {Yang}, \citenamefont {Liu},\ and\ \citenamefont
  {Shao}}]{Yan2004}%
  \BibitemOpen
  \bibfield  {author} {\bibinfo {author} {\bibfnamefont {Y.-a.}\ \bibnamefont
  {Yan}}, \bibinfo {author} {\bibfnamefont {F.}~\bibnamefont {Yang}}, \bibinfo
  {author} {\bibfnamefont {Y.}~\bibnamefont {Liu}}, \ and\ \bibinfo {author}
  {\bibfnamefont {J.}~\bibnamefont {Shao}},\ }\href@noop {} {\bibfield
  {journal} {\bibinfo  {journal} {Chem. Phys. Lett.}\ }\textbf {\bibinfo
  {volume} {395}},\ \bibinfo {pages} {216} (\bibinfo {year}
  {2004})}\BibitemShut {NoStop}%
\bibitem [{\citenamefont {Zhou}\ and\ \citenamefont {Shao}(2008)}]{Shao2008}%
  \BibitemOpen
  \bibfield  {author} {\bibinfo {author} {\bibfnamefont {Y.}~\bibnamefont
  {Zhou}}\ and\ \bibinfo {author} {\bibfnamefont {J.}~\bibnamefont {Shao}},\
  }\href@noop {} {\bibfield  {journal} {\bibinfo  {journal} {J. Chem. Phys.}\
  }\textbf {\bibinfo {volume} {128}},\ \bibinfo {pages} {034106} (\bibinfo
  {year} {2008})}\BibitemShut {NoStop}%
\bibitem [{\citenamefont {Yan}\ and\ \citenamefont {Shao}(2016)}]{Yan2016}%
  \BibitemOpen
  \bibfield  {author} {\bibinfo {author} {\bibfnamefont {Y.-A.}\ \bibnamefont
  {Yan}}\ and\ \bibinfo {author} {\bibfnamefont {J.}~\bibnamefont {Shao}},\
  }\href@noop {} {\bibfield  {journal} {\bibinfo  {journal} {Front. Phys.}\
  }\textbf {\bibinfo {volume} {11}},\ \bibinfo {pages} {110309} (\bibinfo
  {year} {2016})}\BibitemShut {NoStop}%
\bibitem [{\citenamefont {Polyakov}\ and\ \citenamefont
  {Rubtsov}(2018{\natexlab{a}})}]{Polyakov2018b}%
  \BibitemOpen
  \bibfield  {author} {\bibinfo {author} {\bibfnamefont {E.~A.}\ \bibnamefont
  {Polyakov}}\ and\ \bibinfo {author} {\bibfnamefont {A.~N.}\ \bibnamefont
  {Rubtsov}},\ }\href@noop {} {\enquote {\bibinfo {title} {Information loss
  pathways in a nuremically exact simulation of a non-markovian open quantum
  system},}\ } (\bibinfo {year} {2018}{\natexlab{a}}),\ \bibinfo {note} {to be
  published}\BibitemShut {NoStop}%
\bibitem [{\citenamefont {Polyakov}\ and\ \citenamefont
  {Rubtsov}()}]{Polyakov2017b}%
  \BibitemOpen
  \bibfield  {author} {\bibinfo {author} {\bibfnamefont {E.~A.}\ \bibnamefont
  {Polyakov}}\ and\ \bibinfo {author} {\bibfnamefont {A.~N.}\ \bibnamefont
  {Rubtsov}},\ }\href@noop {} {\enquote {\bibinfo {title} {Stochastic dressed
  wavefunction: a numerically exact solver for bosonic impurity model dynamics
  within wide time interval},}\ }\Eprint {http://arxiv.org/abs/1712.04279}
  {arXiv:1712.04279 [cond-mat]} \BibitemShut {NoStop}%
\bibitem [{\citenamefont {Polyakov}\ and\ \citenamefont
  {Rubtsov}(2018{\natexlab{b}})}]{Polyakov2017a}%
  \BibitemOpen
  \bibfield  {author} {\bibinfo {author} {\bibfnamefont {E.~A.}\ \bibnamefont
  {Polyakov}}\ and\ \bibinfo {author} {\bibfnamefont {A.~N.}\ \bibnamefont
  {Rubtsov}},\ }\href@noop {} {\bibfield  {journal} {\bibinfo  {journal} {AIP
  Conf. Proc.}\ }\textbf {\bibinfo {volume} {1936}},\ \bibinfo {pages} {020028}
  (\bibinfo {year} {2018}{\natexlab{b}})}\BibitemShut {NoStop}%
\bibitem [{\citenamefont {Tanimura}(1990)}]{Tanimura1990}%
  \BibitemOpen
  \bibfield  {author} {\bibinfo {author} {\bibfnamefont {Y.}~\bibnamefont
  {Tanimura}},\ }\href@noop {} {\bibfield  {journal} {\bibinfo  {journal}
  {Phys. Rev. A}\ }\textbf {\bibinfo {volume} {41}},\ \bibinfo {pages} {6676}
  (\bibinfo {year} {1990})}\BibitemShut {NoStop}%
\bibitem [{\citenamefont {Breuer}(2007)}]{Breuer2007}%
  \BibitemOpen
  \bibfield  {author} {\bibinfo {author} {\bibfnamefont {H.-P.}\ \bibnamefont
  {Breuer}},\ }\href@noop {} {\bibfield  {journal} {\bibinfo  {journal} {Phys.
  Rev. A}\ }\textbf {\bibinfo {volume} {75}},\ \bibinfo {pages} {022103}
  (\bibinfo {year} {2007})}\BibitemShut {NoStop}%
\bibitem [{Note1()}]{Note1}%
  \BibitemOpen
  \bibinfo {note} {Here $\intop \nolimits D\left [z\right ]$ is understood as
  the limit $\intop \nolimits D\left [z\right ]=\protect \qopname \relax
  m{lim}_{M\to \infty }\protect \frac {1}{\pi ^{M}}\intop \nolimits dz_{x}\left
  (\omega _{1}\right )dz_{y}\left (\omega _{1}\right )\protect \ldots \intop
  \nolimits dz_{x}\left (\omega _{M}\right )dz_{y}\left (\omega _{M}\right )$
  for a discretization of the frequency axis $\omega _{1},\protect \ldots
  \omega _{M}$, and $z_{x}\left (\omega \right ),z_{y}\left (\omega \right )$
  are understood as real and imaginary parts correspondingly of the complex
  number $z\left (\omega \right )$.}\BibitemShut {Stop}%
\bibitem [{\citenamefont {Milonni}(1993)}]{Milonni1993}%
  \BibitemOpen
  \bibfield  {author} {\bibinfo {author} {\bibfnamefont {P.~W.}\ \bibnamefont
  {Milonni}},\ }\href@noop {} {\emph {\bibinfo {title} {The quantum vacuum: an
  introduction to quantum electrodynamics}}}\ (\bibinfo  {publisher} {Academic
  Press},\ \bibinfo {address} {London},\ \bibinfo {year} {1993})\BibitemShut
  {NoStop}%
\bibitem [{\citenamefont {Holstein}(1959)}]{Holstein1959}%
  \BibitemOpen
  \bibfield  {author} {\bibinfo {author} {\bibfnamefont {T.}~\bibnamefont
  {Holstein}},\ }\href@noop {} {\bibfield  {journal} {\bibinfo  {journal} {Ann.
  Phys. (NY)}\ }\textbf {\bibinfo {volume} {8}},\ \bibinfo {pages} {325}
  (\bibinfo {year} {1959})}\BibitemShut {NoStop}%
\end{thebibliography}%

\end{document}